\documentclass[usenatbib]{mnras}
\usepackage{newtxtext,newtxmath}
\usepackage[T1]{fontenc}

\DeclareRobustCommand{\VAN}[3]{#2}
\let\VANthebibliography\thebibliography
\def\thebibliography{\DeclareRobustCommand{\VAN}[3]{##3}\VANthebibliography}

\usepackage{graphicx}	% Including figure files
\usepackage{amsmath}	% Advanced maths commands
\usepackage{caption}
\usepackage{subcaption}
\usepackage{booktabs}
\usepackage{makecell}
\usepackage{array}
\usepackage{soul}
\usepackage{rotating} % Rotating figure and caption
\usepackage{pdflscape}

\title[Tidal features in HSC-SSP]{Analysing the prevalence of tidal features in HSC-SSP using self-supervised representation learning}

\author[A. Desmons et al.]{
A. Desmons,$^{1}$\thanks{E-mail: a.desmons@unsw.edu.au}
S. Brough,$^{1}$
F. Lanusse,$^{2}$
L. Canepa,$^{1}$
A. Khalid$^{1}$
\\
$^{1}$School of Physics, University of New South Wales, NSW 2052, Australia\\
$^{2}$AIM, CEA, CNRS, Universit\'e Paris-Saclay, Universit\'e Paris Diderot, Sorbonne Paris Cit\'e, F-91191 Gif-sur-Yvette, France\\
}

\date{
Accepted 2025 September 19. Received 2025 September 19; in original form 2025 May 5}

\pubyear{2025}

\begin{document}
\label{firstpage}
\pagerange{\pageref{firstpage}--\pageref{lastpage}}
\maketitle

\begin{abstract}
We use a combination of self-supervised machine learning and visual classification to identify tidal features in a sample of 34,331 galaxies with stellar masses $\log_{10}(M_{\star}/\rm{M}_{\odot})\geq9.5$ and redshift $z\leq0.4$, drawn from the Hyper Suprime-Cam Subaru Strategic Program (HSC-SSP) optical imaging survey. We assemble the largest sample of 1646 galaxies with confirmed tidal features, finding a tidal feature fraction $f=0.06^{+0.05}_{-0.01}$. We analyse how the incidences of tidal features and the various classes of tidal features vary with host galaxy stellar mass, photometric redshift, and colour, as well as halo mass. We find an increasing relationship between tidal feature fraction and host galaxy stellar mass, and a decreasing relationship with redshift. We find more tidal features occurring in group environments with $12.0<\log_{10}(M_{200}/\rm{M}_{\odot})<14.0$ than in the field or in denser, cluster environments. We also find that the central galaxies of the most massive  (log$_{10}$($M_{200}$/M$_{\odot}$)~$>$~14.1) groups and clusters exhibit higher rates of tidal features than the satellite galaxies. We find good agreement between the trends we observe and the results obtained from purely visual or other automated methods, confirming the validity of our methodology and that using machine learning can drastically reduce the workload of visual classifiers, having needed to visually classify less than 30 per cent of our sample. Such methods will be instrumental in classifying the millions of suitable galaxies to be observed by large upcoming imaging surveys such as the Vera C. Rubin Observatory’s Legacy Survey of Space and Time. 
\end{abstract}


%%%%%%%%%%%%%%%%%%%%%%%%%%%%%%%%%%%%%%%%%%%%%%%%%%

%%%%%%%%%%%%%%%%% BODY OF PAPER %%%%%%%%%%%%%%%%%%

\section{Introduction}
\label{sec:intro}
\begin{figure*}
    \centering
	\includegraphics[width=0.95\textwidth]{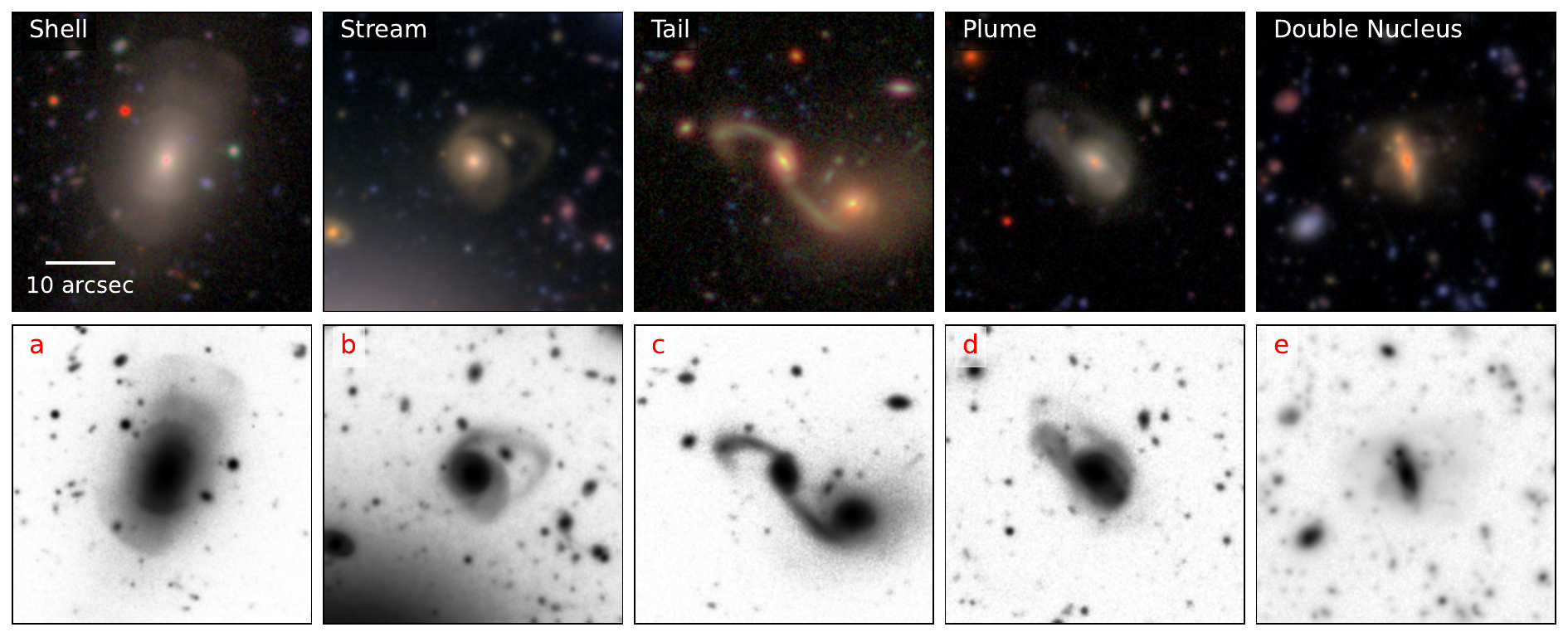}
    \caption{Examples of each type of tidal feature in our classification scheme in panels a-e. The top row shows the arcsinh stretched $g$, $r$, $i$ false colour image from HSC-SSP, the bottom row shows the image generated using the arcsinh-based scaling algorithm introduced by \citet{Gordon2024DecalsTFCNN}.}
    \label{fig:TF_examples}
\end{figure*}

The hierarchical structure formation model of the universe postulates that mergers play an important role in the evolution of galaxies, and that the recent ($z\lesssim2$, e.g. \citealt{Oser2010GalFormationPhases}) growth of the most massive galaxies is dominated by mergers with lower mass galaxies (e.g. \citealt{Lacey1994NBodyMergeRate, Cole2000Hierarchical, Robotham2014GAMAClosePair, Martin2018MergeMorphTransform}). Ongoing and past galaxy mergers leave traces of their interactions in the form of tidal features, which are diffuse, non-uniform regions of stars in the outskirts of galaxies composed of stellar material pulled out from the host or companion galaxy during an interaction (e.g. \citealt{Toomre1972BridgeTails}). Fig.~\ref{fig:TF_examples} shows examples of these tidal features from the optical Hyper Suprime-Cam Subaru Strategic Program (HSC-SSP; \citealt{Aihara2018HSCSurveyDesign}), and demonstrates the variety that exists in tidal feature configuration and morphology. These features are expected to remain observable for $\sim3$~Gyr \citep{Mancillas2019ETGS_merger_hist, Huang2022HSCTidalFeatETG} and their morphology and composition contains information about the merging history of a galaxy, making them an essential tool for the study of galaxy evolution. Shells are circular, often concentric, arcs centered around a galaxy (Fig.~\ref{fig:TF_examples}a). They are thought to form primarily from radial mergers, especially for smaller satellites, or mergers with low angular momentum \citep{Quinn1984ShellsEllipGals, Hendel2015TidalDebOrbit, Pop2018ShellsIllustris}, and can originate from both major and minor mergers (e.g. \citealt{Pop2018ShellsIllustris,KadoFong2018HSCTidalFeat}). Streams (Fig.~\ref{fig:TF_examples}b) are narrow filaments that orbit a host galaxy. They are formed from the tidal disruption of infalling satellites on more circular orbits \citep{Hendel2015TidalDebOrbit} and originate primarily from minor mergers \citep{Mancillas2019ETGS_merger_hist, Sola2022TailsvsStreams}. Tails (Fig.~\ref{fig:TF_examples}c) are also elongated structures that emanate from a galaxy, but unlike streams, they are composed of material expelled from the host galaxy and as a consequence are usually thicker and brighter than streams (e.g. \citealt{Sola2022TailsvsStreams}). Tails are associated with major mergers and tend to have shorter lifetimes than shells or streams \citep{Mancillas2019ETGS_merger_hist}.

Assembling samples of galaxies exhibiting tidal features is difficult due to the very low surface brightness of these features, easily reaching $\mu_r>27~\rm{mag~arcsec}^{-2}$. Such depths are often not achieved by wide-field optical imaging surveys and consequently, many tidal features remain unidentified due to simply not being visible in imaging data. With the next generation of wide-field optical imaging surveys on the horizon, such as the Vera C Rubin Observatory’s Legacy Survey of Space and Time (LSST; \citealt{Ivezic2019LSST}), predicted to reach\footnote{Peter Yaochim, LSST surface brightness limit derivations \url{https://smtn-016.lsst.io/}} $\mu_r\sim30.3~\rm{mag~arcsec}^{-2}$\citep{Martin2022TidalFeatMockIm} and output imaging of millions of galaxies for which tidal features can be seen, assembling large samples of tidal features from which we can draw statistically robust conclusions about the galaxy evolution process is becoming feasible. One major obstacle in assembling such samples is the method of classification, as most past works focusing on the identification and characterisation of tidal features in observational data have relied solely on visual classification to assemble their samples of tidal features (e.g. \citealt{Tal2009EllipGalTidalFeat, Sheen2012PostMergeSigs, Atkinson2013CFHTLSTidal, Hood2018RESOLVETidalFeat, Bilek2020MATLASTidalFeat, Martin2022TidalFeatMockIm, Giri2023ETGRemnants, Rutherford2024MNRAS_SAMI_TF, Yoon2024A_ETG_rot, Yoon2024B_ETG_env}). Classifying billions of galaxy images using visual identification alone is virtually impossible, and even the use of methods relying on large community-based projects such as Galaxy Zoo \citep{Lintott2008GalZoo, Darg2010GalZooFracMarge} and GALAXY CRUISE \citep{Tanaka2023GalCruise} poses numerous challenges, such as ensuring classifications are consistent across all classifiers. Recent years have seen a drastic increase in the use of automated methods to alleviate the workload of visual classifiers (e.g. \citealt{KadoFong2018HSCTidalFeat, Huang2022HSCTidalFeatETG})
and the use of machine learning for the classification of tidal features and merging galaxies (e.g. \citealt{Pearson2019DeepLearnMergers, Snyder2019IllustrisAutoMergerClass, Walmsley2019CNNTidalFeat, Bickley2021CNNTidalIllustris, Dominguez2023MockTidalCNN, Omori2023ML_TF_env, SuelvesPearson2023CNNSDSSMergers, Desmons2024, Ferreira2024MNRASML_UNIONS_mergers, Gordon2024DecalsTFCNN}). Such methods provide an ideal solution to the challenges associated with visual classification by greatly reducing the amount of human input required and the work required from visual classifiers. However, these techniques typically require a trade-off of sample purity for classification efficiency, a factor that must be considered with the science aim of the work in mind, when deciding to use visual or automated methods. 

In this work, we aim to identify and classify tidal features in one of the largest samples of HSC-SSP galaxies analysed to-date, consisting of 34,331 galaxies, to analyse how the incidence of tidal features varies with stellar mass, photometric redshift, colour, and halo mass. Our priority is to ensure that the sample of galaxies with tidal features we assemble is highly pure to maximize the accuracy of our results and conclusions. The large size of the sample and the requirement that it be highly pure means that solely visual or automated methods are not ideal, as due to the low number of galaxies with tidal features compared to the global galaxy population even a machine learning model with a low false positive rate will have a significant impact on the purity of the tidal feature sample (e.g. \citealt{Huertas-Company2023DLAstroReview}). Instead, we choose to use a combination of machine learning and visual classification to classify our sample, where the machine learning significantly reduces the amount of galaxies that must be visually classified, and the visual classification enables us to assemble a sample whose high purity could not be achieved through use of our chosen machine learning model alone. To reduce the visual classification workload and identify tidal feature candidates we use the self-supervised machine learning model trained by \citet{Desmons2024} which assigns high scores to galaxies likely to possess tidal features. We then visually classify the highest-ranked candidates to create a pure sample of galaxies with tidal features, and classify the features into various classes including shells, stellar streams, tidal tails, and asymmetric haloes. We examine how the trends observed between tidal feature incidence and host galaxy properties using this partially automated method compare with the results obtained from purely visual or other automated methods.

Section~\ref{sec:Methods} details our sample selection and methods, including details of the machine learning model and our visual classification scheme. In Section~\ref{sec:results} we present our results, including a calculation of the tidal feature fraction for our full sample, and an analysis of how this tidal feature fraction and the incidence of the various classes of tidal features vary with host galaxy stellar mass, halo mass, photometric redshift, and colour. Section~\ref{sec:disc} presents a comparison of our results with those in the literature, both from observational and simulation-based works, and works that use purely observational methods, or automated methods that differ from those we use here. Our conclusions are presented in Section~\ref{sec:conc}. Throughout this paper, we assume a flat $\Lambda$CDM cosmology with $h=0.7$, $H_0=100~h~\rm{km}~\rm{s}^{-1}~\rm{Mpc}^{-1}$, $\Omega_m=0.3$,and $\Omega_{\Lambda}=0.7$.

\section{Methods}
\label{sec:Methods}
\subsection{Data sources and sample selection}
\label{sec:data}
In this work we use galaxy images and data sourced from the HSC-SSP Public Data Release 2 (PDR2; \citealt{Aihara2019HSCSecondData}). The HSC-SSP survey \citep{Aihara2018HSCSurveyDesign} is a three-layered, \textit{grizy}-band imaging survey carried out with the Hyper Suprime-Cam on the 8.2m Subaru Telescope located in Hawaii. Although the HSC-SSP Public Data Release 3 (\citealt{Aihara2022HSCThirdData}) is currently available and is an updated version of the HSC-SSP PDR2, we do not use this release due to the differences in the data treatment pipelines. HSC-SSP PDR2 has been widely tested for low surface brightness studies (e.g. \citealt{Huang2018HSCStellarHalo,Huang2020WeakLens,Li2022MassiveGalOutskirt,MartinezLombilla2023GAMAIntragroupLight}) and fulfils the requirements for our study. The HSC-SSP survey comprises three layers: Wide, Deep, and Ultradeep, which are observed to varying surface brightness depths. We use the Deep/Ultradeep (D/UD) fields, which span an area of $27$ and $3.5$~deg$^{2}$, respectively, and reach $\mu_{r}\sim29.82$~mag arcsec$^{-2}$ ($3\sigma,~10''\times10''$; \citealt{MartinezLombilla2023GAMAIntragroupLight}), a surface brightness depth faint enough to detect tidal features. We use HSC-SSP data not only for its depth, but also because it is the data used to train the \citet{Desmons2024} machine learning model we use to detect tidal features. Using data from the same source as the model training set ensures that we already know how well the model performs on these data and prevents us having to retrain the model. The HSC-SSP PDR2 has a median $\it{i}$-band seeing of 0.6 arcsec and a spatial resolution of 0.168 arcsec per pixel.

Our work includes an analysis of tidal feature incidence as a function of galaxy stellar mass, photometric redshift, and $g-i$ galaxy colour. We obtain stellar masses and  photometric redshifts from the \texttt{Mizuki} catalogue available from the HSC-SSP database. These properties are inferred from the \texttt{MIZUKI} \citep{Tanaka2015PhotoZBayesionPrior} template-fitting code, which generates templates using stellar population synthesis models. To obtain $g-i$ colours for our sample we initially chose to use rest-frame magnitudes provided in the \texttt{Mizuki} catalogue. However, upon examination of the resulting colour-stellar mass diagram, the data showed clear discretization in the rest-frame colour distribution, leading to concerns about the quality of the template matching. We therefore decide to use the \texttt{cmodel} $g$- and $i$-band apparent magnitudes provided in the HSC-SSP forced photometry catalogues. When using these to calculate the $g-i$ colours of galaxies, we apply k-corrections to each band using the methods outlined in \citet{Chilingarian2010MNRASkcorrect1} and \citet{Chilingarian2012MNRASkcorrect2}. To limit the effects of systematics, such as template-matching issues, on our results we verified that our results were consistent when using different galaxy property catalogues. One of these was the \texttt{DEmP} catalogue \citep{Hsieh&Yee2014DEmP}, available from the HSC-SSP database, which provides stellar masses and photometric redshifts. The other included calculating stellar masses using the empirical relation between stellar mass, $g-i$ colour, and $i-$band luminosity presented in \citet{Taylor2011GAMAMassEst}. All results presented in Section~\ref{sec:results} remain qualitatively unchanged regardless of the data used.

\begin{figure}
    \centering
	\includegraphics[width=0.99\columnwidth]{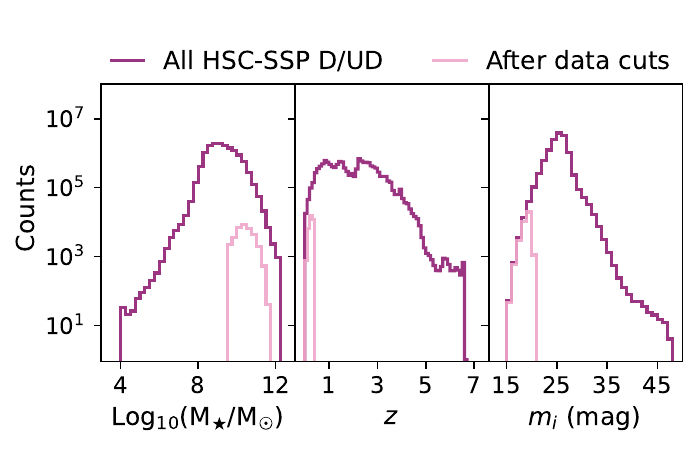}
    \caption{Distribution of galaxy properties in the HSC-SSP Deep/UltraDeep (D/UD) fields (purple) and our sample after applying the stellar mass, redshift, and $i$-band apparent magnitude cuts (pink).}
    \label{fig:pre_post_data_cuts}
\end{figure}

To assemble our sample of galaxies we apply a series of data cuts to the full sample of HSC-SSP PDR2 D/UD galaxies. We set $i$-band magnitude limits $15\leq{i}\leq20$~mag, a photometric redshift limit $z\leq0.4$, and a stellar mass limit $\log_{10}(M_{\star}/\mathrm{M}_{\odot})\geq9.5$. These limits ensure that we only select galaxies for which tidal features should be visible. \citet{Martin2022TidalFeatMockIm} find that at a surface brightness limit equivalent to our data of $\mu_{r}\sim30~\rm{mag}~\rm{arcsec}^{-2}$, the fraction of flux detectable in tidal features reaches zero at redshift $z=0.4$, and that beyond this redshift the visibility of tidal features becomes significantly impacted by the point spread function of the survey.  In their analysis of HSC-SSP Wide galaxies, Kado-Fong et al. (2018) find very few observable tidal features around galaxies with $\log_{10}(M_{\star}/\mathrm{M}_{\odot})<9.5$, and although their surface brightness limit ($\mu_{r}\sim26.4~\rm{mag}~\rm{arcsec}^{-2}$) is shallower than ours, we use this as our absolute stellar mass limit to ensure that we can detect tidal features. The distributions of galaxy properties for the full HSC-SSP D/UD fields and for our sample after applying these data cuts are shown in Fig.~\ref{fig:pre_post_data_cuts}. Additionally, we select only images which have at least three exposures in each band ($g,~r,~i,~z,~y$) to ensure galaxies have full depth and colour information. We also remove objects which have been flagged for any reason including objects with \texttt{cmodel} fit failures and bad photometry, objects affected by bright sources, or objects with failures relating to radius measurement. After applying these initial cuts we obtain a sample of 39,309 galaxies. 

We access the HSC-SSP galaxy images using the \textsc{Unagi} \textsc{Python} tool \citep{Huang2019Unagi} which, given a galaxy’s right ascension and declination, allows us to create multi-band ‘HSC cutout’ images of size 256~$\times$~256 pixels ( 42~$\times$~42 arcsecs), centred around each galaxy. Each cutout is downloaded in five ($g,~r,~i,~z,~y$) bands.

 Several objects in HSC-SSP are affected by bad segmentation and `bright galaxy shredding' where bright ($i<19$ mag) galaxies are deblended into multiple objects \citep{Aihara2018HSCSurveyDesign}. Galaxy shredding affects as much as 15 per cent of bright HSC-SSP galaxies, particularly late-type galaxies \citep{Aihara2018HSCSurveyDesign}. To ensure that we do not have objects affected by bad segmentation or shredding in our dataset, we run the \texttt{find\_galaxy} function from the MGE fitting method and software developed by \citet{Cappellari2002MGE}. This algorithm is applied to the 256~$\times$~256 pixel cutouts to identify galaxies in our images. We then remove galaxies which have no valid fit, or galaxies for which the centre of the fit is not within 3 pixels of the image centre, from our sample. This leaves us with a final sample composed of 34,331 galaxies with median $i$-band magnitude $i=19.16\pm0.73$~mag, photometric redshift $z=0.26\pm0.08$, and stellar mass $\log_{10}(M_{\star}/\mathrm{M}_{\odot})=10.38\pm0.41$.

To further improve the robustness of our conclusions and limit the impact of the large uncertainties associated with photometric redshifts we repeat the analysis presented in Section~\ref{sec:results} on a subset of galaxies in our sample for which spectroscopic redshifts are available. We do this by cross-matching our sample of 34,331 galaxies with data from two spectroscopic surveys. The first of these is the Galaxy and Mass Assembly (GAMA; \citealt{Driver2011GAMADataRel}) survey, a multi-wavelength spectroscopic survey carried out using the optical AAOmega multi-object spectrograph (\citealt{Hopkins2013GAMASpectra, Liske2015GAMADR2}). The second set of spectroscopic data is sourced from the Early Data Release of the Dark Energy Spectroscopic Instrument (DESI EDR; \citealt{DESICollab_Overview2022, DESICollab_EDR2024}). DESI is a 5000-fiber multiobject spectrograph installed on the Mayall 4-meter telescope at Kitt Peak National Observatory \citep{DESICollab_target_design2016, DESICollab_inst_design2016} and the EDR contains spectroscopic redshifts for 1.8 million targets observed during Survey Validation. In our sample of 34,331 galaxies, we identify 13,977 galaxies with DESI EDR redshifts, and 4472 galaxies with GAMA redshifts, resulting in 18,469 galaxies for which we have spectroscopic redshifts, or $54$ per cent of our sample. The spectroscopic subsample has median $i$-band magnitude $i=18.89\pm0.70$~mag, photometric redshift $z=0.24\pm0.08$, and stellar mass log$_{10}$($M_{\star}$/M$_{\odot})=10.41\pm0.43$. While the galaxies in this spectroscopically confirmed sample are slightly brighter and have slightly lower redshifts, these are within $1\sigma$ of the overall sample distribution and the spectroscopic subset covers the full range of brightness, redshift, and stellar masses present in our parent sample. We repeat the analysis presented in Section~\ref{sec:results} on this subset of galaxies with spectroscopic redshifts and find that the conclusions presented in Section~\ref{sec:results} also hold true qualitatively for this spectroscopic subset.  

\begin{figure*}
    \centering
	\includegraphics[width=0.95\textwidth]{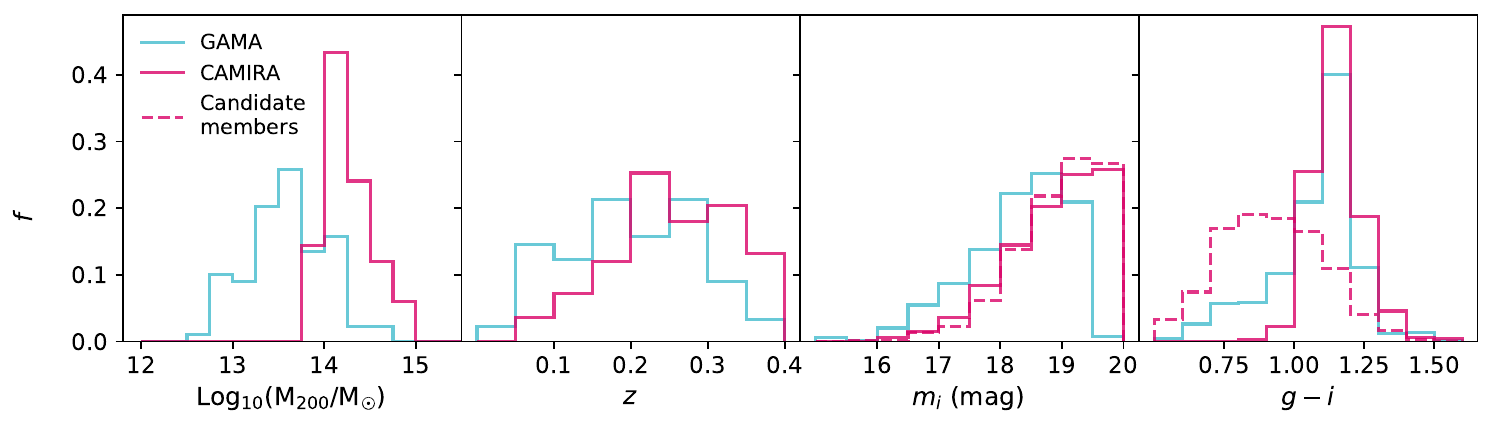}
    \caption{Distribution of group, cluster, or individual galaxy properties for the final GAMA (blue), CAMIRA (solid pink), and candidate member (dashed pink) samples that will be considered in our analysis. From left to right, distribution of: group/cluster halo mass, group/cluster BCG redshift (spectroscopic for GAMA, photometric when spectroscopic is not available for CAMIRA), member galaxy $i$-band magnitude, member galaxy k-corrected $g-i$ colour.}
    \label{fig:GAMA_CAM_dists}
\end{figure*}

\subsubsection{Cluster and group catalogues}
\label{sec:Data_clusters}
This work also includes an analysis of the prevalence of tidal features with respect to environment, using halo mass as a proxy for environment. We obtain halo masses by identifying galaxies in our sample that are in groups or clusters using two different catalogues. The first of these is CAMIRA \citep{Oguri2018CAMIRA}, an optically-selected cluster catalogue of HSC-SSP galaxies. This cluster-finding algorithm, detailed in \citet{Oguri2014CAMIRA_alg}, relies on the expectation that all massive clusters possess a `red sequence' of early-type galaxies (e.g. \citealt{Gladders_Yee2000ClusterRedSeq}) and uses a stellar population synthesis model to identify these likely red sequence galaxies and group them into clusters. From the full CAMIRA cluster catalogue, we keep only galaxies which are in our main sample of $\sim34,000$ galaxies, by applying the same series of data cuts detailed in Section~\ref{sec:data}. Our final CAMIRA sample consists of 2142 cluster members from 97 clusters. To obtain halo masses for these clusters we use the weak-lensing calibrated richness-mass relation for CAMIRA clusters presented in \citet{Murata2019ClusterMassRich}:

\begin{align}
    \label{eq:halo_mass}
    \it{B}~\rm{ln}\left(\frac{\it{M_{\rm{200}}}}{\it{M}_{\rm{pivot}}}\right) =& ~ln~\it{N}~-~\it{A}~-~B_{z}\rm{ln}\left(\frac{1~+~\it{z}}{1~+~z_{\rm{pivot}}}\right)\nonumber\\
    &-
    \it{C_{z}}\left[\rm{ln}\left(\frac{1~+~\it{z}}{1~+~z_{\rm{pivot}}}\right)\right]^{2}
\end{align}

where $M_{200}$ is the halo mass of the cluster, $z$ is the cluster redshift, and $N$ is the cluster richness. The richness parameter, provided in the CAMIRA catalogue, is defined as the number of red member galaxies within an aperture of radius $\rm{R}\sim1$~$h^{-1}~$Mpc, with stellar masses $M_*~\geq~10^{10.2}~M_{\odot}$ \citep{Oguri2014CAMIRA_alg}. For the definition of the other parameters we refer the reader to \citet{Murata2019ClusterMassRich}. We use the parameters estimated using \textit{WMAP} cosmology, which have values $M_{\rm{pivot}}=3\times10^{14}~h^{-1}~M_{\odot}$, $z_{\rm{pivot}}=0.6$, $A=3.36$, $B=0.83$, $B_z=-0.20$, and $C_z=3.51$.

The CAMIRA catalogue provides a list of the red sequence members for each cluster but does not identify any non-red sequence members. As we are interested in tidal feature fractions, we want to make sure our sample is not biased to red members only.  We therefore assemble our own list of additional candidate members for each cluster using their right ascension, declination, and redshift. The CAMIRA catalogue provides the coordinates of the Brightest Cluster Galaxy (BCG) of each cluster, identified by the algorithm. For each cluster we calculate the distance of the confirmed red-sequence CAMIRA member farthest from the cluster BCG. We then find galaxies in the main sample which are closer to the cluster BCG than this maximum distance and whose photometric redshifts are within 1$\sigma$ of the cluster's, where the 1$\sigma$ accuracy of the HSC-SSP photometric redshifts is given as $0.05(1+z_{\rm{phot}})$ \citep{Tanaka2018HSCSSP_photoz}. For the 97 CAMIRA clusters with members in the main sample, we identify 1223 additional candidate members.

In addition to CAMIRA clusters, we also use the GAMA survey and its associated group catalogue \citep{Robotham2011GAMAGroups} to identify galaxies in our main sample that belong to lower mass group systems. The GAMA Galaxy Group Catalogue (G$^3$C; \citealt{Robotham2011GAMAGroups}) uses a friends-of-friends (FoF) algorithm to identify galaxy groups by creating links between galaxies based on their projected and radial separations. We make use of two of the four data tables that make up the G$^3$C, namely the \texttt{G3CGalv10} and \texttt{G3CFoFGroupv10} catalogues. We only consider groups with at least 5 members and retain only galaxies which are present in our main sample by matching their right ascension and declination within 1 arcsecond. This results in a sample of 95 groups consisting of 525 galaxies, for which we obtain spectroscopic redshifts from the \texttt{G3CGalv10} catalogue. We remove 2 groups (4 galaxies) from this sample which have BCG spectroscopic redshift $z_{\rm{spec}}>0.4$, resulting in a sample of 93 groups consisting of 521 galaxies. To ensure that the halo masses for the GAMA groups are comparable to those obtained for the CAMIRA clusters, we calculate halo masses for the GAMA groups using the weak-lensing calibrated luminosity-mass relation for GAMA groups presented in \citet{Viola2015GAMA_WL_group_masses}:
\begin{equation}
    \frac{M_{200}}{10^{14}h^{-1}M_{\odot}} = 0.95\left(\frac{L_{\rm{grp}}}{10^{11.5}h^{-2}L_{\odot}}\right)^{1.16}
\end{equation}
where $M_{200}$ is the halo mass of the group, and $L_{\rm{grp}}$ is the $r$-band luminosity of the GAMA group, provided in the \texttt{G3CFoFGroupv10} catalogue.

As a final step, we check for overlap between the 97 CAMIRA clusters and 93 GAMA groups with members in our main sample and find 105 galaxies that are shared between 16 CAMIRA clusters and 23 GAMA groups. Of the 16 CAMIRA clusters, 7 are composed of multiple GAMA groups and require us to choose whether to use the CAMIRA or GAMA grouping. We perform a case by case evaluation of these clusters by looking at the position of the galaxies in projected separation, using right ascension, declination, and the redshifts of the galaxies. For 5 of these CAMIRA clusters, we use the GAMA groups they are composed of, and for the remaining 2 clusters we use the CAMIRA definition of each cluster. The remaining 9 CAMIRA clusters only overlap with single GAMA groups. For these we use the GAMA definition of the groups due to the greater reliability of the GAMA spectroscopy.

Our final sample of group and cluster galaxies consists of 492 GAMA members belonging to 89 distinct GAMA groups, and 1871 CAMIRA members and 1062 candidates belonging to 83 distinct CAMIRA clusters. Fig.~\ref{fig:GAMA_CAM_dists} shows the distributions of halo mass and redshift for the groups and clusters these members belong to, as well as the members' $i$-band magnitude and $g-i$ colour distributions. The GAMA groups and CAMIRA clusters share similar redshift distributions but the GAMA groups exhibit a wider range of halo masses than the CAMIRA clusters which have a minimum halo mass $\log_{10}(M_{200}/\rm{M}_{\odot})\geq13.8$. Due to the fainter depths achieved by the HSC-SSP survey, the CAMIRA clusters contain fainter member galaxies, and the red sequence method used to identify the CAMIRA cluster members means that the range of $g-i$ colours of CAMIRA cluster members is more concentrated toward redder galaxies, while the additional candidate member galaxies we identify are more concentrated toward bluer galaxies.

When investigating the relationship between tidal feature prevalence and environment we also consider a subsample of field galaxies alongside our subsample of galaxies in groups and clusters. We define this sample of field galaxies as all galaxies in our sample which have not been assigned to a group or cluster. We can be sure that these galaxies do not belong to clusters with halo masses greater than $\log_{10}(M_{200}/\rm{M}_{\odot})\sim14.0$, as the CAMIRA catalogue identifies clusters across the area covered by the entire HSC-SSP survey. However, there is a chance that they could belong to groups with lower halo masses than this, as the GAMA catalogue does not cover the same area as our HSC-SSP sample. Measurements using this field subsample are likely a good indication of the tidal feature trends for field galaxies, but this limitation should be considered. To ensure that our conclusions are not affected by differing galaxy property distributions within these subsamples we verify that they share similar redshift and stellar mass distributions. The CAMIRA + candidate members, GAMA, and field subsamples have mean stellar masses $\log_{10}(M_{\star}/\mathrm{M}_{\odot})=10.43\pm0.39$, $10.49\pm0.43$, and $10.39\pm0.41$, and mean redshifts $z=0.26\pm0.07$, $0.21\pm0.07$, and $0.26\pm0.08$, respectively.

In Section~\ref{sec:halo_dep} we consider central and satellite galaxies in the halo separately. However, the CAMIRA cluster catalogue does not provide central galaxies as a separate entity to the BCGs. When conducting this analysis, we therefore take the CAMIRA-defined BCG of each cluster as the central galaxy. The GAMA group catalogue does provide information about the central galaxy of each cluster, independent from the BCG, in this case we use the GAMA definition of the central galaxy. Of the 47 GAMA central galaxies present in our cluster sample, 38 (80 per cent) of them are also the BCGs of their respective groups. All galaxies in our group and cluster sample that are not BCGs in CAMIRA or central galaxies in GAMA are considered satellite galaxies. 

Further details of our cluster candidate member selection and our final GAMA and CAMIRA group and cluster selections are provided in Appendix~\ref{sec:app_cluster}, along with an investigation of the effects of these selections on our results.

\subsection{Image pre-processing and model architecture}
\label{sec:mod_arch}
To partially automate the detection of tidal features in our sample we use the pre-trained model designed by \citet{Desmons2024}. This model is composed of a self-supervised encoder which outputs lower-dimensional encodings of the input images, and a linear classifier which takes these encodings as input and outputs a number between 0 and 1, where numbers closer to 1 indicate a higher probability of tidal features being present.

The encoder section of the model is composed of a ResNet-20 \citep{He2016deep} network followed by two fully connected layers of size 128, each followed by a batch-normalisation layer. Prior to being encoded, a series of random augmentations are applied to the images, including addition of Gaussian noise, image rotation, and random cropping to 96~$\times$~96 pixels. These augmentations are embedded into the model and cannot be removed but do not affect model performance as the model encodings were trained to be invariant under these data transformations. The encoder takes in 128~$\times$~128 pixel images and outputs 128-dimensional representations, which are then passed on to the classifier. The classifier is a very simple model composed only of a  fully connected layer with a sigmoid activation, which outputs a single number between 0 and 1. Outputs closer to 1 indicate a higher likelihood of tidal features being present in the image.
We do not perform training from scratch but instead use the pre-trained versions of encoder and classifier presented in \citet{Desmons2024}. Both models were trained on images drawn from the D/UD layers of the HSC-SSP PDR2. The encoder was trained using $\sim44,000$ unlabelled images, while the classifier was trained using 300 images of galaxies without tidal features and 300 images of galaxies possessing tidal features drawn mainly from a sample assembled by \citet{Desmons2023GAMA}. For further detail on the architecture of the network, the performance of the model, and the training datasets we refer the reader to \citet{Desmons2024}.

Before images are passed on to the model, we apply a pre-processing function to normalize them. This is done by taking a random subsample of 1000 galaxies from our dataset and calculating the median absolute deviation $\sigma_{\text{pixel count}}$ for each band $(g,r,i,z,y)$ of this subsample. The entire sample is then normalized by dividing each image band by the corresponding $3\sigma_{\text{pixel count}}$ and then taking the hyperbolic sine of this. The image input size for the \citet{Desmons2024} model is 128~$\times$~128 pixels meaning our downloaded 256~$\times$~256 pixel images must be resized. Instead of cropping the images, we choose to use interpolation of pixel fluxes to resize the images. This ensures that galaxy outskirts and potential tidal features are not cropped out for galaxies which occupy a larger portion of the images.

\subsection{Detection and classification}
\label{sec:detect}

Our aim for this work is to analyse how the incidence of tidal features varies with stellar mass, photometric redshift, colour, and halo mass. Hence, one of our priorities is to ensure that the tidal feature sample we assemble is highly pure, with minimal contamination from non-tidal galaxies, to maximize the accuracy of our results and conclusions. This is why, instead of simply using the outputs of the machine learning model and a threshold to separate galaxies with and without tidal features, we opt to use a combination of model output and visual classification. The combination of these two methods greatly reduces the amount of galaxies in our sample that must be visually classified, while also ensuring that our tidal feature sample is uncontaminated. When performing visual classification of the tidal features present in our galaxy sample we follow a similar classification scheme to those used in \citet{Bilek2020MATLASTidalFeat}, \citet{Martin2022TidalFeatMockIm}, \citet{Desmons2023GAMA}, and \citet{Khalid2024TF_Sims}. The tidal feature categories include:
\begin{itemize}
    \item Shells: Concentric radial arcs or ring-like structures around a galaxy.
    \item Tidal tails: Prominent, elongated structures pulled out from the host galaxy. These usually have similar colours to that of the host galaxy.
    \item Stellar streams: Narrow filaments orbiting a host galaxy. These differ from tails in that they do not consist of material unbound from the host galaxy but instead consist of stars stripped from a smaller companion galaxy.
    \item Plumes or asymmetric stellar haloes: Diffuse features in the outskirts of the host galaxy, lacking well-defined structure like stellar streams or tails, or galaxies where the structure of the stellar halo is clearly asymmetric.
    \item Double nuclei: Galaxies which are visibly merging but where both objects are still clearly separated. To belong to this class, objects must exhibit clear signs of interaction (i.e. they cannot only be close pairs).
\end{itemize}

Examples of each type of tidal feature are shown in Fig.~\ref{fig:TF_examples}. These categories are not mutually exclusive as galaxies can have multiple types of features and hence can belong to multiple categories. When performing the visual classification we do not count the incidence of each type of tidal features, only whether they are present. The classification was performed by the first author. We acknowledge that having a sole classifier could introduce a level of bias to the classifications (e.g. \citealt{Martin2022TidalFeatMockIm, Sola2025Strrings_streams}). While having several classifiers repeating the classifications could help quantify and reduce this intrinsic noise, we opt not to do this due the significant additional visual classification workload this would introduce.

Since this work includes an analysis of tidal features present in groups and clusters, we must consider the impact that the intracluster light (ICL), a diffuse, low surface brightness component extending through the cluster but not bound to a particular galaxy (e.g. \citealt{Mihos2005ICL}), may have on our tidal feature classification. As a significant portion of the ICL is considered to originate from stars ripped out of galaxies during interactions (e.g. \citealt{Contini2021ICLReview, Montes2022ICLReview}), at some point during the interaction those stars will likely take the form of tidal features. If the diffuse light clearly originates from an ongoing interaction between two galaxies in the cluster, we classify this as a tidal feature. However, smoothly distributed diffuse light surrounding a central galaxy but not clearly originating from it would not be considered as a tidal feature in our classification scheme. With respect to how the light constituting the ICL may impact the visibility of tidal features, \citet{Khalid2024TF_Sims} conducted an analysis of the effect of background contamination on the detectability of tidal features. Although this analysis was conducted using a simple toy model, their results suggest that the surface brightness limit of the survey is likely to impact tidal feature detectability before the light from the ICL does.

\begin{figure}
    \centering
	\includegraphics[width=0.99\columnwidth]{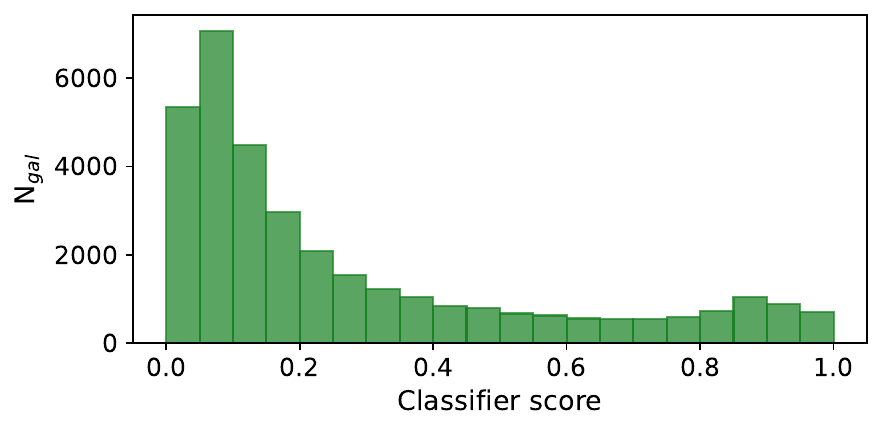}
    \caption{Distribution of the classifier scores assigned to all 34,331 galaxies in our sample.}
    \label{fig:Class_scores_dist}
\end{figure}

\begin{figure}
    \centering
	\includegraphics[width=0.99\columnwidth]{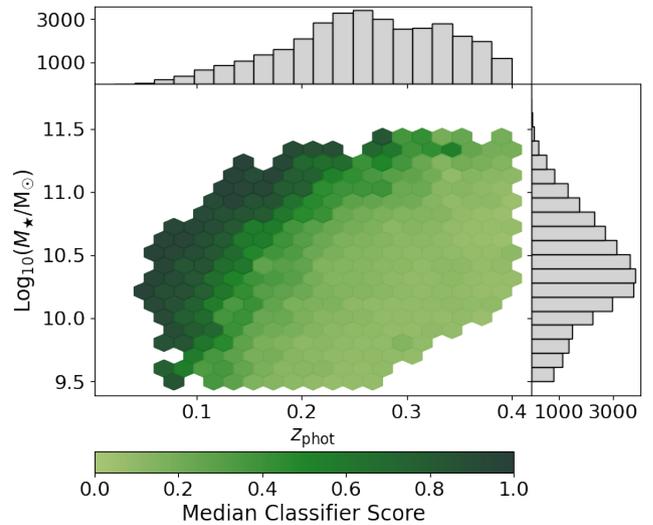}
    \caption{The central figure shows the photometric redshift and stellar mass distribution of our sample in hexagonal bins. Each bin is coloured according to its median classifier score with values as shown in the bottom colour bar. Only bins with at least 10 galaxies are shown. The top and right panels show the 1D histograms over redshift and mass for our full sample.}
    \label{fig:Mass_z_score}
\end{figure}

We start by applying the model to all of the galaxies in our sample in order to obtain a classifier score between 0 and 1 for each galaxy. The result of the model application is shown in Fig.~\ref{fig:Class_scores_dist} where, as expected, the majority of galaxies are predicted to be non-tidal with scores close to zero. Fig.~\ref{fig:Mass_z_score} shows the stellar mass and photometric redshift distribution of our sample, coloured according to the median classifier score in each hexagonal bin. From this figure it is evident that galaxies assigned high classifier scores are concentrated in the low redshift and high stellar mass regions. This is not surprising as tidal features are more likely to be visible for brighter, and hence more massive, lower redshift galaxies (e.g. \citealt{KadoFong2018HSCTidalFeat,Martin2022TidalFeatMockIm}). However, this effect can also be in part attributed to a bias the model has toward assigning high classifier scores to galaxies with these properties. This bias was discussed in \citet{Desmons2024} and is why, in order to ensure that our tidal feature sample is pure, we also combine model outputs with visual classification. 

\begin{figure}
    \centering
	\includegraphics[width=0.99\columnwidth]{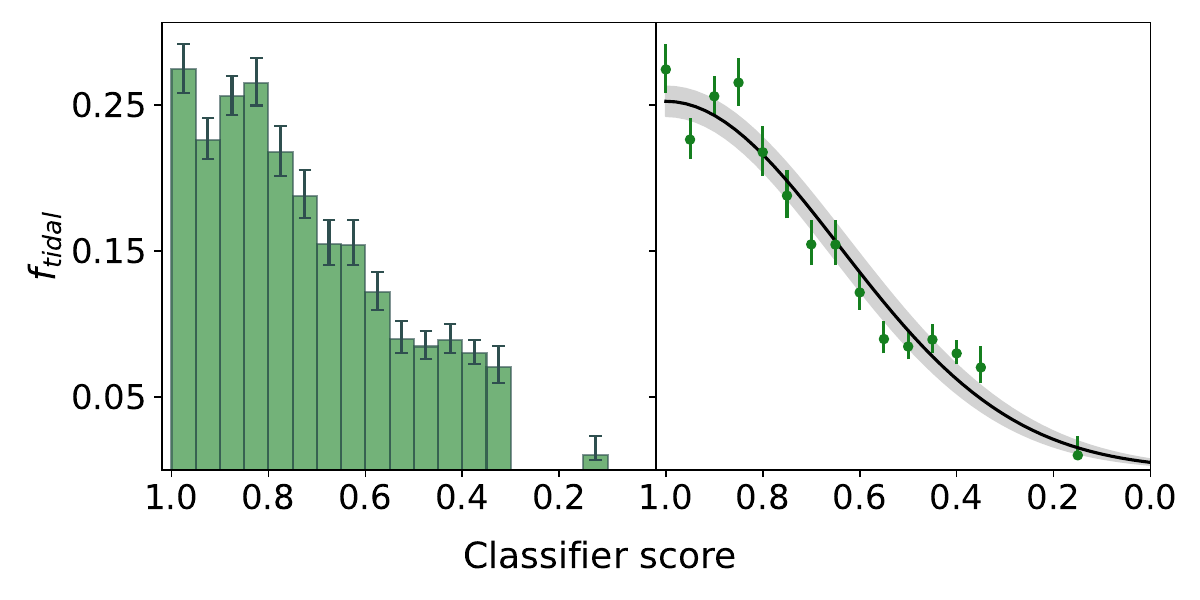}
    \caption{Left panel: results of our visual classification showing the fraction of tidal features in classifier score bins for the 10,000 galaxies with the highest classifier scores. The rightmost bin shows the tidal feature fraction for an extra 200 galaxies that were classified, with classifier scores $\sim0.1$. Right panel: The black line shows the half-normal fit to the results of our visual classification, the grey areas show the 1$\sigma$ uncertainty to the fit.}
    \label{fig:Vis_class_fit}
\end{figure}

Having obtained classifier scores for all the galaxies in our sample, we visually classify the galaxies in order of descending classifier score, starting with those assigned the highest classifier scores. We continue the visual classification until the incidence of tidal features plateaus. The left panel in Fig.~\ref{fig:Vis_class_fit} displays the results of our visual classification, showing that the tidal feature fraction decreases as a function of classifier score, and plateaus in the five bins between classifier scores 0.55 to 0.3. These five bins have a median tidal feature incidence of 8.27 per cent. This is where we stop our visual classification, having classified 10,000 galaxies out of our total 34,331 galaxies, consistent with classifier scores above 0.35. Out of these 10,000 galaxies, we visually identify 1646 galaxies with tidal features. Fig.~\ref{fig:Vis_class_fit} highlights the importance of our visual classification in increasing the purity of our sample of galaxies with tidal features. If we had instead used a purely machine learning-based approach with a conservative threshold of 0.8, our sample of galaxies with tidal features would have been significantly contaminated, with 75 per cent of the sample being galaxies without tidal features.

Our work will include calculating an overall tidal feature fraction, and seeing how this fraction varies with stellar mass, halo mass, photometric redshift, and colour. In order to draw statistically robust conclusions about the trends we observe between the tidal feature fraction and host galaxy properties we also want to constrain this tidal feature fraction by calculating lower and upper bounds. We know that our sample cannot have fewer galaxies with tidal features than those confirmed by visual classification. Hence, we can calculate a lower bound on the number of tidal features present in our sample by assuming that the only tidal features in our sample are those that are identified by visual classification. Given the decreasing trend of tidal feature incidence with decreasing classifier scores, we can be fairly certain that the fraction of tidal features for classifier scores below 0.3 does not exceed the stable level of 8.27 per cent reached between classifier scores 0.55 to 0.3. Hence, we can use this knowledge to calculate an upper bound on the number of tidal features present in our sample below classifier scores of 0.3 by assuming a maximum tidal feature incidence rate of 8.27 per cent below these scores. To obtain a middle or `true' value for the incidence of tidal features below classifier scores of 0.3 we want to get an estimate of the amount of tidal features that are present in galaxies that were not visually classified. To get a more robust picture of the amount of tidal features in these galaxies we visually classify an extra 200 galaxies with classifier scores $\sim0.1$, shown in the rightmost bin of the left panel of Fig.~\ref{fig:Vis_class_fit}. We then fit a half normal distribution to the results of our visual classification, as shown in the right panel of Fig.~\ref{fig:Vis_class_fit}. The equation for our fit is: 
\begin{equation}
        f = \rm{A}e^{\frac{-(1-x)^2}{\rm{B}}}
	\label{eq:fit_eq}
\end{equation}

where $f$ is the fraction of tidal features in a classifier score bin given the upper bound $x$ of the bin, and the fit parameters are $\rm{A}=0.253$ and $\rm{B}=0.257$ for classifier score bins of width 0.05. When calculating the middle value for the tidal feature fraction we use our visual classification results for galaxies with classifier scores above 0.3, and use this fit to estimate the amount of tidal features for classifier scores below 0.3, using classifier score bins of width 0.05. We can express the tidal feature calculation using the equation:                                                                          
\begin{equation}
\label{eq:tf_frac}
    f = \frac{N_{\rm{tidal}}}{N_{\rm{gal}}} = \frac{N_{\rm{tidal,vis}}+N_{\rm{tidal,est}}}{N_{\rm{gal}}}
\end{equation}
where $N_{\rm{gal}}$ is the total number of galaxies in our sample, 34,311 galaxies, and $N_{\rm{tidal}}$ is the number of galaxies with tidal features in our full sample. $N_{\rm{tidal, vis}}$ is the number of visually-classified tidal features and $N_{\rm{tidal, est}}$ is the estimated number tidal features in galaxies that were not visually-classified, which varies depending on whether we are calculating the lower or upper bound, or the middle estimate.
when calculating the:
\begin{itemize}
    \item Lower bound: $N_{\rm{tidal, est}}$ = 0.
    \item Middle point: $N_{\rm{tidal, est}}$ is the estimated number of tidal features remaining for classifier scores below 0.3, according to the half-normal fit defined in Equation~\ref{eq:fit_eq}.
    \item Upper bound: $N_{\rm{tidal, est}}$ is the number of tidal features remaining if we estimate a stable tidal feature incidence rate of 8.27 per cent below classifier scores of 0.3.
\end{itemize}

In this work, the use of the machine learning model allows us to analyse a large sample of galaxies in a short time and drastically reduces the number of galaxies which need to be visually classified to less than a third of our full sample. The use of visual classification allows us to create a pure sample of galaxies with tidal features, and to better estimate the tidal feature fraction and constrain it with lower and upper bounds.

\begin{figure}
    \centering
	\includegraphics[width=0.95\columnwidth]{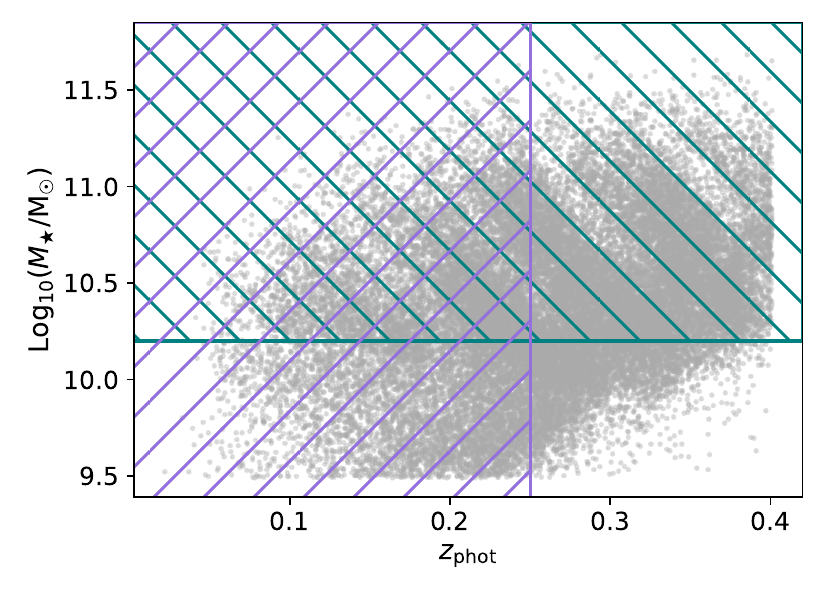}
    \caption{The grey points show the photometric redshift and stellar mass distribution of our full sample. The teal hatching indicates the area covered by the mass cut subsample, where a lower stellar mass limit $\log_{10}(M_{\star}/\mathrm{M}_{\odot})\geq10.2$ is imposed. The purple hatching indicates the area covered by the redshift cut subsample, where an upper photometric redshift limit $z\leq0.25$ is imposed.}
    \label{fig:Mass_z_cuts}
\end{figure}

\begin{figure*}
    \centering
	\includegraphics[width=0.95\textwidth]{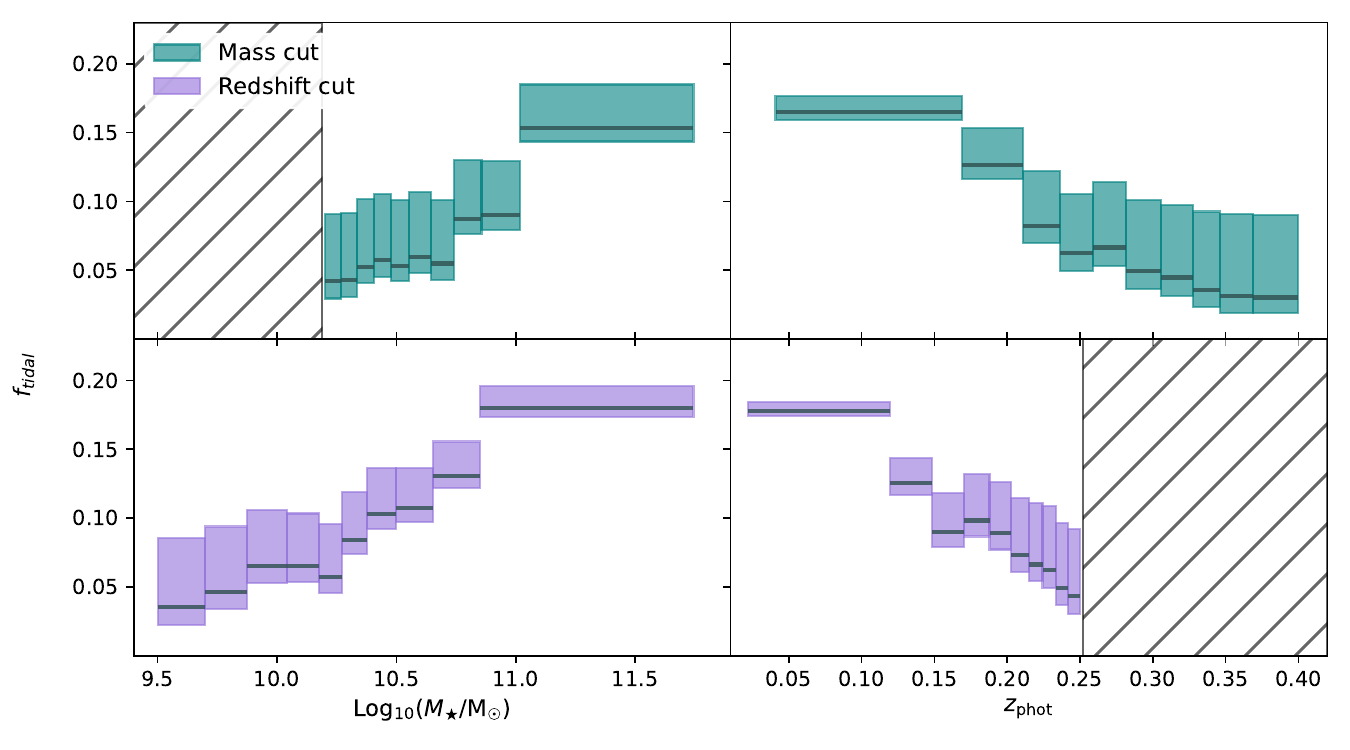}
    \caption{The tidal feature fraction as a function of stellar mass (left panels) and photometric redshift (right panels) for the mass-cut subsample (top panels) and redshift-cut subsample (bottom panels). The subsamples are binned such that each bin contains an equal number of galaxies, the bins in the mass-cut subsample each contain $\sim2340$ galaxies, and the bins in the redshift-cut subsample each contain $\sim1480$ galaxies. The left and right edges of each box show the extent of the mass or redshift bin, the bottom and top edges of the boxes show the lower and upper bound on the tidal feature fraction, and the central line in each box shows the middle estimate. For both the mass-cut and redshift-cut subsamples, the tidal feature fraction increases with increasing stellar mass and decreases with increasing redshift.}
    \label{fig:Mass_z_box_plots}
\end{figure*}

\section{Results}
\label{sec:results}
In this section we first provide a value for the overall tidal feature fraction, including lower and upper bounds on this quantity. We then analyse the dependence of this tidal feature fraction on stellar mass, photometric redshift, and host galaxy colour. We also examine the relationship between the presence of tidal features and the environment of their host galaxies by looking at whether the tidal feature fraction varies with cluster halo mass. We look at whether there exists a relationship between the types of tidal features present and the stellar mass, redshift, and colour of their host galaxies. 

We start by calculating the overall tidal feature occurrence fraction in our full sample, according to Equation~\ref{eq:tf_frac}. We obtain an overall tidal feature fraction of $f_{\rm{middle}}=0.06$ with $f_{\rm{lower}}=0.05$ and $f_{\rm{upper}}=0.11$.

\subsection{Dependence on stellar mass and redshift}
\label{sec:Mass_z_dep}

When analysing the stellar mass and redshift dependence of the tidal feature fraction we consider two separate volume-limited subsamples. The first subsample is mass-limited with a lower stellar mass limit $\log_{10}(M_{\star}/\mathrm{M}_{\odot})\geq10.2$, and contains 23,412 galaxies. The second subsample is redshift limited with an upper photometric redshift limit $z\leq0.25$, and contains 14,782 galaxies. These stellar mass and redshift limits are chosen empirically, based on where the stellar mass-redshift distribution is visibly impacted by low completeness. The area covered by these two subsamples in the stellar mass-redshift space is shown in Fig.~\ref{fig:Mass_z_cuts}.

The calculated fractions of tidal features as a function of stellar mass and photometric redshifts for both of these subsamples are presented in Fig.~\ref{fig:Mass_z_box_plots}. The fraction of tidal features increases continuously with stellar mass for both subsamples. For the redshift-cut subsample, which allows us to examine the full mass range, the tidal feature fraction increases from 0.04~(0.02,~0.09) for $9.50\leq\log_{10}(M_{\star}/\mathrm{M}_{\odot})<9.70$ to 0.18~(0.16,~0.20) for $10.85\leq\log_{10}(M_{\star}/\mathrm{M}_{\odot})\leq11.74$, where the numbers in parentheses show the lower and upper bounds for the relevant fractions. The tidal feature fraction decreases continuously with increasing redshift in both subsamples. For the mass-cut subsample, which allows us to examine the full redshift range, the tidal feature fraction decreases from 0.17~(0.15,~0.18) for $0.04\leq$~$z<0.17$ to 0.03~(0.02,~0.10) for $0.37\leq$~$z\leq0.40$. 

\begin{figure}
    \centering
	\includegraphics[width=0.99\columnwidth]{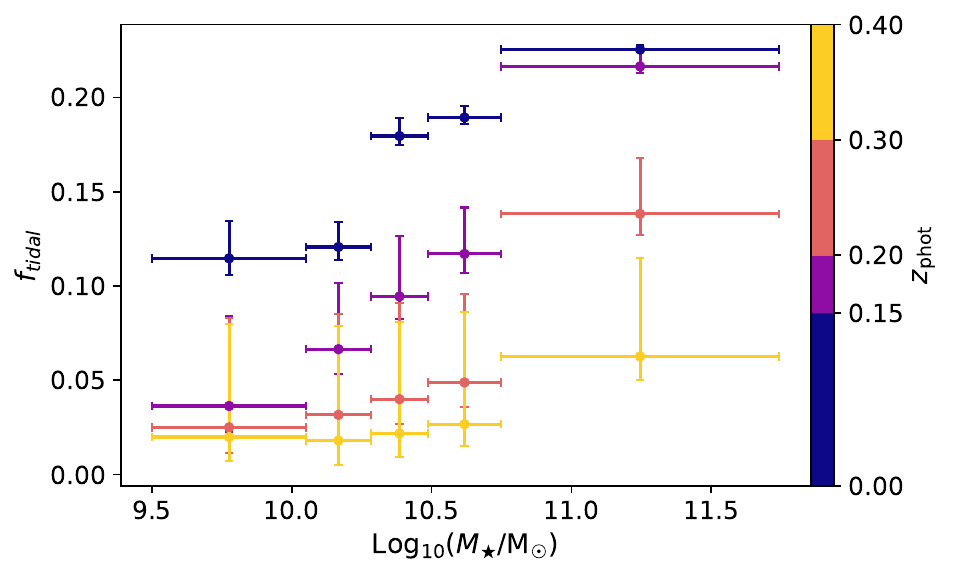}
    \caption{The tidal feature fraction in bins of stellar mass sand photometric redshift. The error bars along the x-axis represent the extent of the stellar mass bins, while the colours indicate the redshift bins. The full sample is binned equally in stellar mass such that each mass bin contains $\sim7000$ galaxies but individual mass/redshift bins have varying numbers of galaxies. Each point shows the tidal feature fraction for the corresponding bin, calculated using the `middle point' defined in Section~\ref{sec:detect}, while the upper and lower error bars show the upper and lower bounds for each bin. Beyond redshift $z\sim0.3$ the trend with stellar mass reduces in strength, and the tidal feature fraction only increases in the highest stellar mass bin.}
    \label{fig:TF_frac_mass_z_bins}
\end{figure}

We explore the relationship between tidal feature incidence, stellar mass, and photometric redshift in more detail in Fig.~\ref{fig:TF_frac_mass_z_bins}. This figure shows the fraction of tidal features present in bins of both stellar mass and photometric redshift, allowing us to examine the relationship between tidal features, stellar mass, and redshift simultaneously. The lowest redshift bin covers a wider redshift range, to $z=0.15$, to ensure that each bin contains sufficient galaxies to draw reliable conclusions, as there are significantly fewer galaxies in the range $0\leq$~$z<0.1$. As expected from the results in Fig.~\ref{fig:Mass_z_box_plots}, the tidal feature fraction peaks in the highest stellar mass, lowest redshift bin, and decreases with increasing redshift and decreasing stellar mass. The tidal feature fraction remains highest, within lower and upper bounds, in the highest stellar mass bin for any given redshift bin, and in the lowest redshift bin for any stellar mass bin. As redshift increases, the trend with stellar mass gradually disappears. Above redshift $z=0.3$ the tidal feature fraction remains approximately constant at $f_{\rm{tidal}}\sim0.03$, apart from the highest stellar mass bin, $\log_{10}(M_{\star}/\mathrm{M}_{\odot})\geq10.75$.

\subsection{Dependence on galaxy colour}
\label{sec:col_dep}
\begin{figure}
    \centering
	\includegraphics[width=0.99\columnwidth]{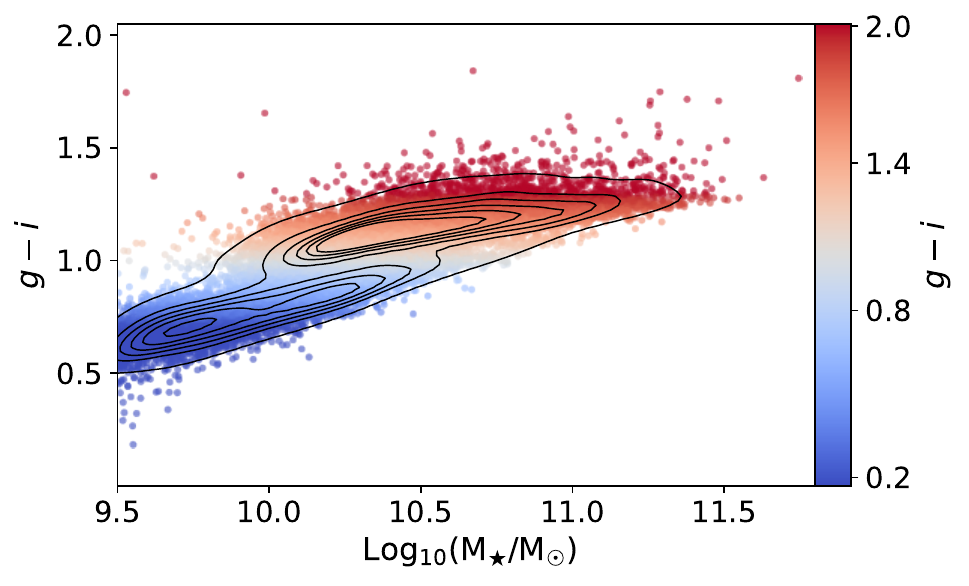}
    \caption{Distribution of stellar mass and $g-i$ colour for the redshift-limited subsample. The solid black lines show the $g-i$ and stellar mass contours representing 5 to 100 per cent of the data across 7 equally spaced levels.}
    \label{fig:g_i_mass_dist}
\end{figure}

In this section we investigate whether tidal features are observed at higher rates for redder or bluer galaxies. To do this, we use the k-corrected $g-i$ colour of the galaxies. When performing our colour analysis, we use the redshift-limited subsample defined in Fig.~\ref{fig:Mass_z_cuts} with the upper photometric redshift limit $z<0.25$. We do this to ensure that all conclusions made here are based on a complete sample in stellar mass. Fig.~\ref{fig:g_i_mass_dist} shows the distribution of stellar mass and $g-i$ colour for our redshift-limited subsample. From this figure it is evident that there exists a strong correlation between stellar mass and $g-i$ colour, and given the relationship between tidal feature fraction and stellar mass, this could impact any relationship we observe between $g-i$ colour and tidal features. To verify whether this is the case we conduct a partial correlation analysis, which tells us whether a relationship between $g-i$ colour and tidal feature occurrence is present once the stellar mass relationship is removed. We obtain partial correlation coefficients $r=0.01~[-0.01,~0.04]$ and $r=0.02~[0.0,~0.05]$ for the Pearson and Spearman methods, respectively, where the numbers in brackets show the 95 per cent parametric confidence intervals around $r$. This indicates that although tidal features in our sample are observed at higher rates around redder galaxies, this is not due to the colour of the galaxies themselves, but rather a consequence of redder galaxies having higher stellar masses than bluer galaxies in our sample.

\subsection{Dependence on halo mass}
\label{sec:halo_dep}

\begin{figure}
    \centering
	\includegraphics[width=0.99\columnwidth]{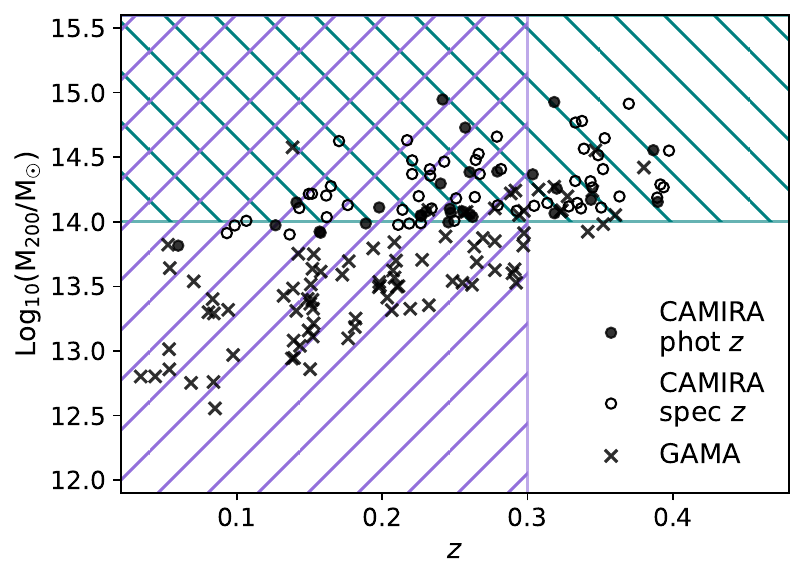}
    \caption{The circle points show the photometric (filled) or spectroscopic (open) redshift and halo mass distribution of the CAMIRA cluster BCGs. The crosses show the spectroscopic redshift and halo mass distribution of the GAMA group BCGs. The teal hatching indicates the area covered by the halo mass-cut subsample, where a lower halo mass limit $\log_{10}(M_{200}/\mathrm{M}_{\odot})\geq14.0$ is imposed. The purple hatching indicates the area covered by the redshift-cut subsample, where an upper redshift limit $z\leq0.3$ is imposed.}
    \label{fig:Halo_mass_z_cuts}
\end{figure}

\begin{figure*}
    \centering
    \includegraphics[width=0.99\textwidth]{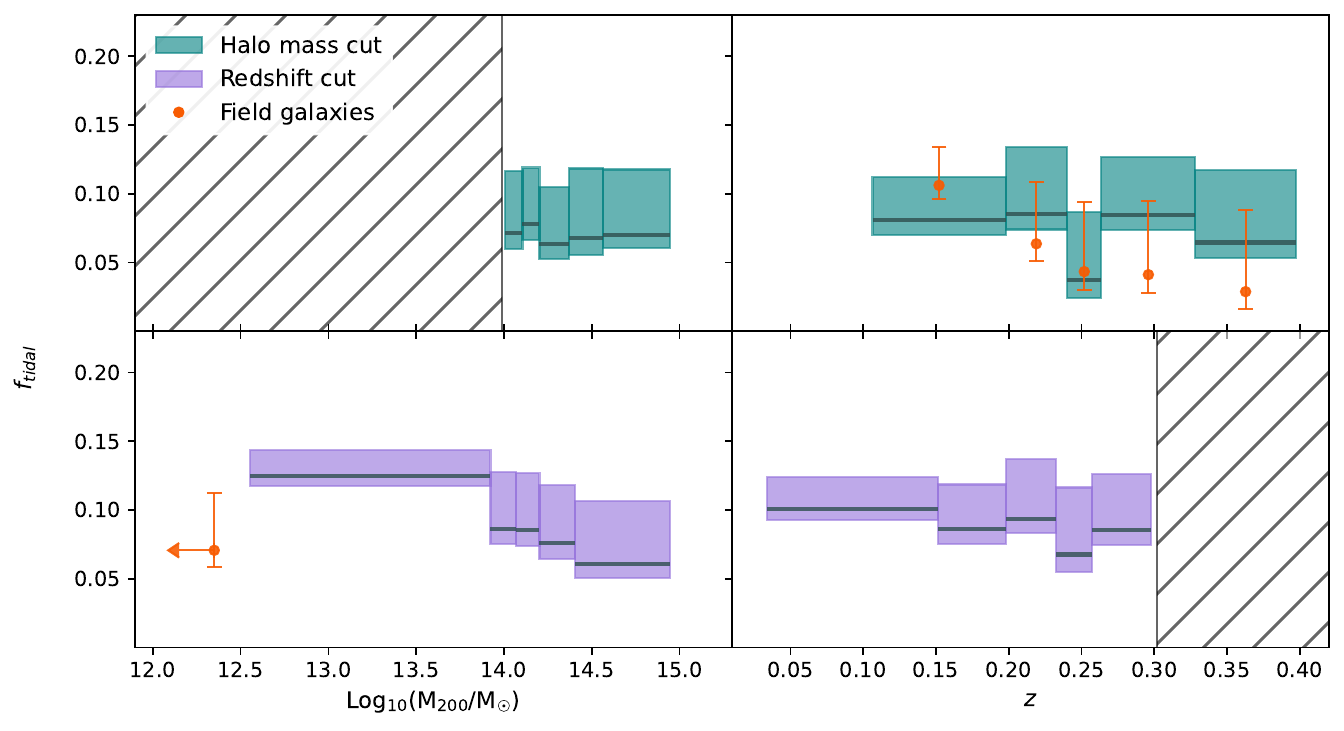}
    \caption{The tidal feature fraction as a function of halo mass (left panels) and BCG redshift (right panels; spectroscopic when available, photometric otherwise) for the halo mass-cut subsample (top panels) and redshift-cut subsample (bottom panels). The subsamples are binned such that each bin contains a similar number of galaxies, the bins in the mass-cut subsample each contain $\sim700$ galaxies, and the bins in the redshift-cut subsample each contain $\sim800$ galaxies. The left and right edges of each box show the extent of the halo mass or redshift bin, the bottom and top edges of the boxes show the lower and upper bound on the tidal feature fraction, and the central line in each box shows the middle estimate. The orange points show the tidal feature fractions, with lower and upper bounds, for galaxies not assigned to groups or clusters (field). For the top right panel this is the tidal feature fraction as a function of galaxy photometric redshift. The arrow on the orange point in the lower left panel indicates that this measurement applies to all galaxies in systems with $\log_{10}(M_{200}/\mathrm{M}_{\odot})<12$, as halo mass measurements are not available for field galaxies. For the redshift-cut subsample, the tidal feature fraction shows a slight decrease with increasing halo mass.}
    \label{fig:Halo_mass_box_plots}
\end{figure*}

In this section we examine whether there exists a relationship between the presence of tidal features and the environment their host galaxies occupy. We do this by limiting our analysis to galaxies in groups and clusters and looking at how the tidal feature fraction varies as a function of the halo mass of these groups and clusters. Higher halo masses generally indicate denser environments, and trends between tidal feature incidence and halo mass can reveal in which environments tidal features are more likely to form or remain visible longer. We use the group and cluster subsample described in Section~\ref{sec:Data_clusters} consisting of 492 GAMA group members, 1871 CAMIRA cluster members, and 1062 CAMIRA cluster candidate members. We follow a similar approach to that in Section~\ref{sec:Mass_z_dep} by considering two separate volume-limited subsamples. The first subsample is halo mass-limited with a lower halo mass limit $\log_{10}(M_{200}/\mathrm{M}_{\odot})\geq14.0$, and contains 2540 galaxies from 89 groups and clusters. The second subsample is redshift limited, based on the redshift of the cluster BCG, with an upper redshift limit $z\leq0.3$, and contains 2772 galaxies from 133 groups and clusters. When setting a BCG redshift limit we use the spectroscopic redshift of the BCG when available, otherwise we use the HSC-SSP photometric redshift. For GAMA, all 89 groups have available spectroscopic redshifts; for CAMIRA, 55 of the 83 clusters have available spectroscopic BCG redshifts. The location of each of these groups and clusters in halo mass-redshift space, along with the area covered by the two volume-limited subsamples, is shown in Fig.~\ref{fig:Halo_mass_z_cuts}.

Fig.~\ref{fig:Halo_mass_box_plots} shows the tidal feature fraction as a function of halo mass and BCG redshift for these two subsamples as well as for field galaxies, as defined in Section~\ref{sec:Data_clusters}. In the redshift-cut subsample in the lower panels, we observe a decrease in the tidal feature fraction with increasing halo mass for galaxies in groups and clusters.  We also see a suggestion that galaxies in groups with halo masses $12.6<\log_{10}(M_{200}/\mathrm{M}_{\odot})<13.9$ have higher rates of tidal features than field galaxies. With regards to cluster redshift, we do not see any trend with the tidal feature fraction in either the halo mass-cut or redshift-cut subsample. Field galaxies do show a slight decrease in tidal feature fraction with increasing redshift.

\begin{figure*}
    \centering
	\includegraphics[width=0.99\textwidth]{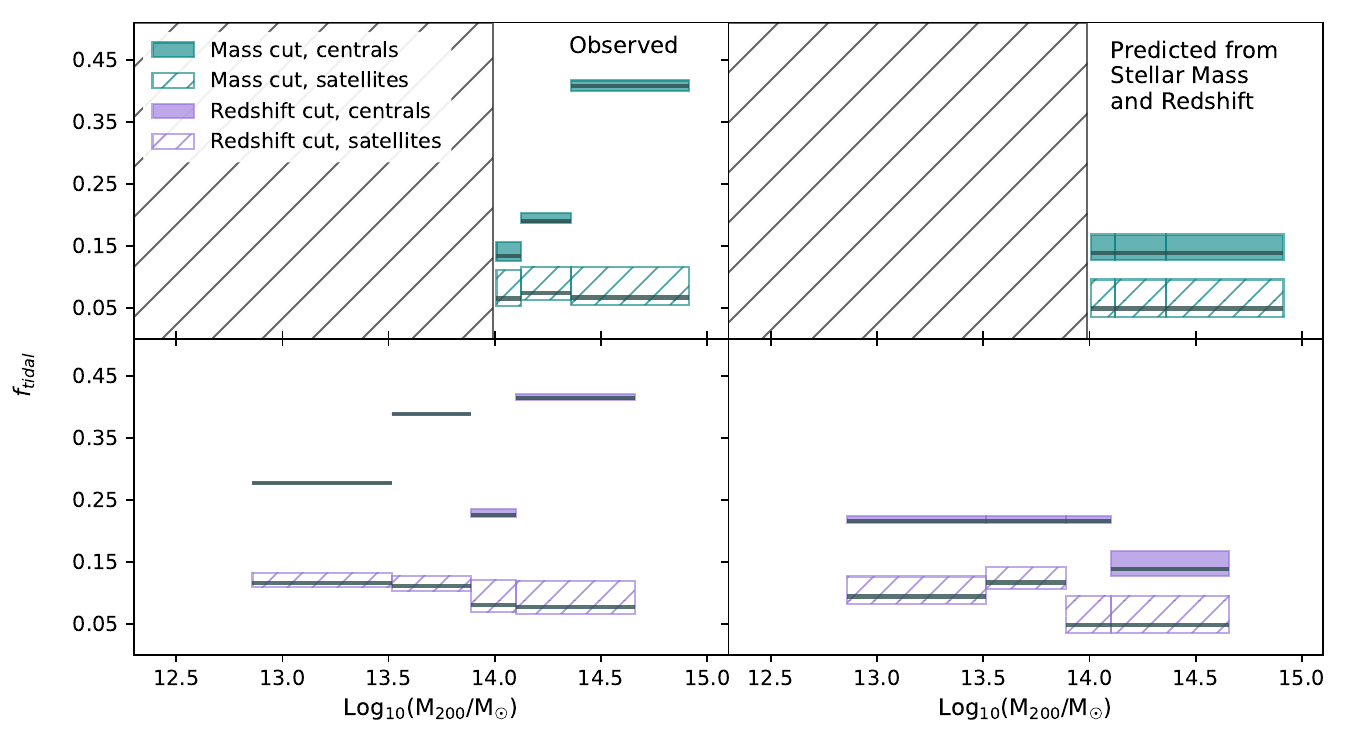}
    \caption{The tidal feature fraction as a function of halo mass for the central and satellite galaxy populations, for the mass-cut subsample (top panels) and redshift-cut subsample (bottom panels). 
    The subsamples are binned such that each centrals bin contains an equal number of galaxies ($\sim25$ galaxies). The solid boxes show the tidal feature fraction for central galaxies, the hatched boxes show the tidal feature fraction for the satellite galaxies, using the same bins as for the centrals. The left panels show the observed tidal feature fraction as a function of halo mass. The right panels show the tidal feature fraction estimated from the tidal feature fraction-stellar mass-redshift relation shown in Fig.~\ref{fig:TF_frac_mass_z_bins}. The left and right edges of each box show the extent of the halo mass bin, the bottom and top edges of the boxes show the lower and upper bound on the tidal feature fraction, and the central line in each box shows the middle estimate.}
    \label{fig:Centrals_sats_box_plots}
\end{figure*}

We look at the trend between tidal features and environment in more detail by considering central and satellite galaxies. The left panels of Fig.~\ref{fig:Centrals_sats_box_plots} show the evolution of the tidal feature fraction as a function of halo mass for centrals and satellites, for both our mass-cut and redshift-cut subsamples. The satellites follow the same trend observed in the left panels of Fig.~\ref{fig:Halo_mass_box_plots}, showing a slight decrease in tidal feature fraction with increasing halo mass in the redshift-cut subsample. This is expected as satellite galaxies are much more numerous than central galaxies, causing them to dominate the trend presented in Fig.~\ref{fig:Halo_mass_box_plots}. The centrals show a different relationship with halo mass, with the tidal feature fraction generally increasing with increasing halo mass, apart for $13.9<\mathrm{log}_{10}(M_{200}/\mathrm{M}_{\odot})\leq14.1$ in the redshift-cut subsample. Most notably, for any given halo mass bin, the centrals have a significantly higher tidal feature fraction than the satellites, with the largest gap being 0.34 (or a tidal feature fraction $\sim6$ times greater) for the highest halo mass bin in the halo mass-cut subsample. In order to make any statements about the environmental dependence of the tidal feature fraction we need to ensure that the trends we see with halo mass cannot be explained by the trends we observed of the tidal feature fraction with galaxy stellar mass and redshift, particularly given that central galaxies are generally more massive than satellite galaxies. To do this we use the same halo mass bins as the left panels of Fig.~\ref{fig:Centrals_sats_box_plots} and calculate the mean stellar mass and photometric redshift of the galaxies in each bin. We then refer to Fig.~\ref{fig:TF_frac_mass_z_bins} to find the expected tidal feature fraction in each bin based on this mean stellar mass and photometric redshift, and plot this fraction the right panels of Fig.~\ref{fig:Centrals_sats_box_plots}. If the trends between tidal feature fraction and halo mass were caused by the trends with stellar mass and redshift, then the plots in the left panels of Fig.~\ref{fig:Centrals_sats_box_plots} should match those in the right panels. For satellite galaxies, the observed tidal feature fractions are generally very slightly higher than those predicted by stellar mass and redshift (apart from in the lowest halo mass bin in the redshift-cut subsample) but all values agree within bounds and the same trend can be observed in both the left and right panels. This suggests that the trend observed with halo mass, both for the overall population (left panels of Fig.~\ref{fig:Halo_mass_box_plots}) and the satellite population is a consequence of the trends observed between tidal feature fraction and stellar mass and redshift. 

The observed tidal feature fraction for the central galaxies is generally higher than the one predicted from the stellar mass and redshift relation, except for the first bin in the halo mass-cut subsample, where the predicted tidal feature fraction is higher than the observed one. For centrals in the halo mass-cut subsample, the relationship between tidal feature fraction and halo mass is different for the observed and predicted panels. The observed panel (left) shows a strong increase in the tidal feature fraction in the highest halo mass bin, whereas the predicted panel (right) shows no increase in this bin. In the redshift-cut subsample, the predicted panel shows an overall stable tidal feature fraction with increasing halo mass with a decrease in the highest halo mass bin, while the observed panel shows a general decrease with a drop at $13.9<\mathrm{log}_{10}(M_{200}/\mathrm{M}_{\odot})\leq14.1$. These opposing trends suggest that stellar mass and redshift are not solely responsible for the varying tidal feature fractions, and location within the cluster environment could play a role. Generally, while stellar mass or redshifts trends partially account for the tidal feature fractions being higher in central galaxies than satellite galaxies, they do not account for the full difference. This suggests that central galaxies in groups and clusters host more tidal features independently of their stellar mass, particularly central galaxies in high halo mass systems ($\log_{10}(M_{200}/\mathrm{M}_{\odot})>14.1$).

\begin{figure}
    \centering
	\includegraphics[width=0.99\columnwidth]{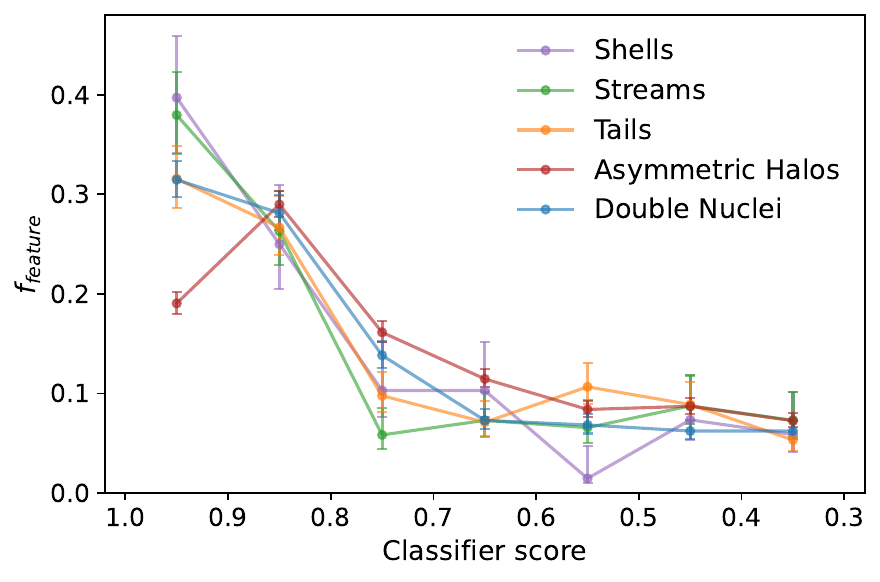}
    \caption{The distribution of classifier scores for each category of tidal features, in classifier score bins of width 0.1. Here, $f_{\rm{feature}}$ is the fraction of galaxies in the specific tidal feature category whose classifier scores are in a given classifier score bin.}
    \label{fig:feat_type_scores}
\end{figure}

\begin{table*}
\caption{Fraction and mean stellar mass, photometric redshift, and $g - i$ colour of host galaxies in 7 categories: galaxies without tidal features, galaxies with any type of feature, and galaxies with shells, tails, streams, asymmetric haloes or plumes, and double nuclei. The fractions are given with their upper and lower bounds. The mean stellar mass and photometric redshift for each subsample is calculated using only the properties of host galaxies with visually-identified features. Uncertainties on these are given to 1$\sigma$.}
\label{tab:feat_types}
\normalsize
\centering
\begin{tabular}{p{2.4cm}>
{\centering}m{2.1cm}>
{\centering}m{2.cm}>
{\centering\arraybackslash}m{1.5cm}}
\toprule
 & {$f$} & {$\overline{\rm{log}_{10}(M_{\star}/\rm{M}_{\odot})}$} & {$\overline{z}$} \\
\midrule
No features & -- & 10.38 $\pm$ 0.40 & 0.26 $\pm$ 0.07 \\
\\
All features & $0.060^{0.047}_{0.012}$ & 10.62 $\pm$ 0.44 & 0.21 $\pm$ 0.08 \\
\\
Shells & $0.0025^{0.0019}_{0.0005}$ & 10.75 $\pm$ 0.49 & 0.17 $\pm$ 0.07 \\
\\
Tails & $0.0082^{0.0064}_{0.0016}$ & 10.55 $\pm$ 0.45 & 0.20 $\pm$ 0.08 \\
\\
Streams & $0.0050^{0.0039}_{0.0010}$ & 10.64 $\pm$ 0.41 & 0.19 $\pm$ 0.08 \\
\\
Asymmetric Haloes & $0.045^{0.035}_{0.009}$ & 10.62 $\pm$ 0.44 & 0.21 $\pm$ 0.08 \\
\\
Double Nuclei & $0.024^{0.019}_{0.005}$ & 10.67 $\pm$ 0.43 & 0.23 $\pm$ 0.08 \\
\\
\bottomrule
\end{tabular}
\end{table*}

\subsection{Tidal feature types and host galaxy properties}
\label{sec:type_dep}

\begin{figure*}
    \centering
	\includegraphics[width=0.99\textwidth]{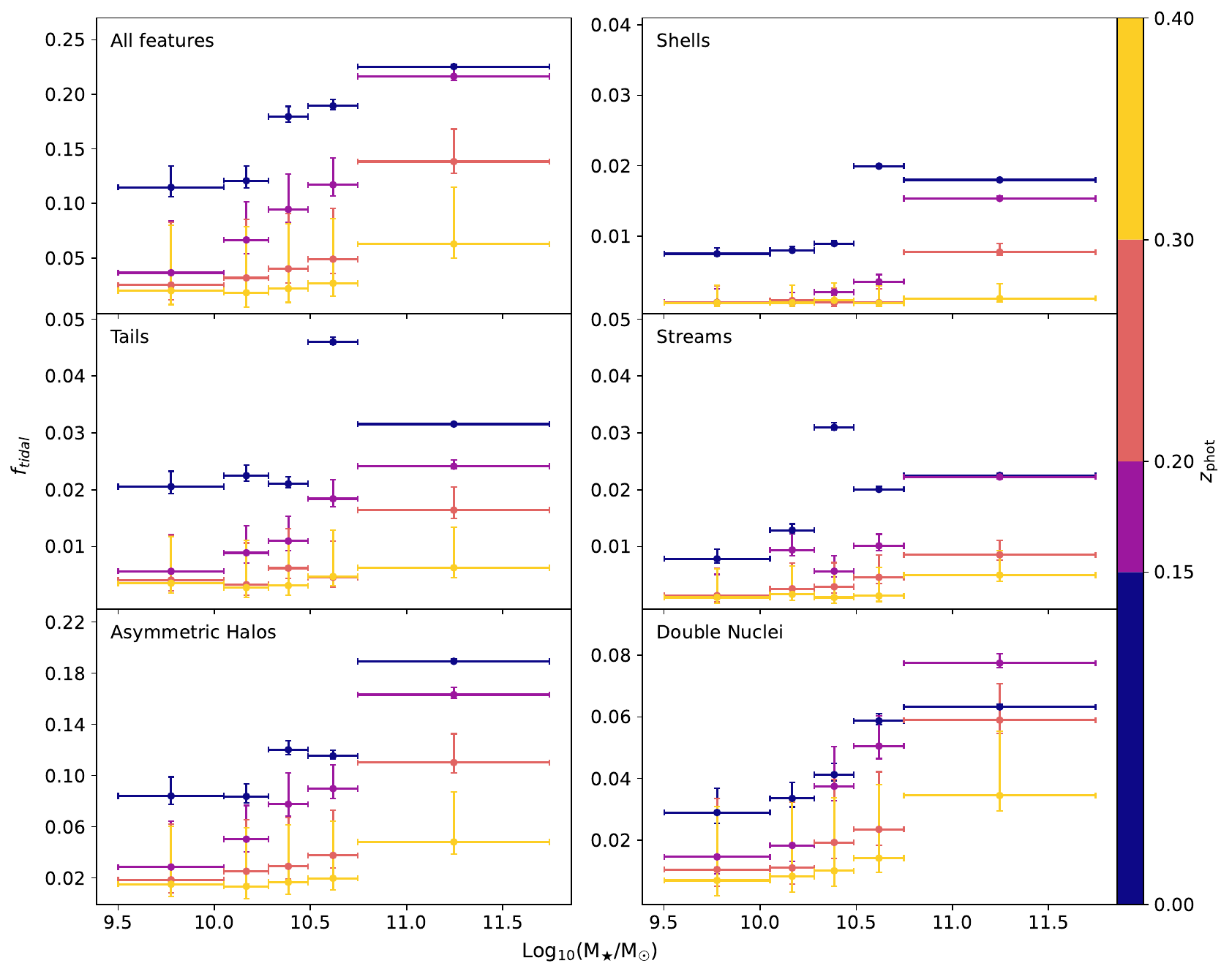}
    \caption{The fraction of all tidal features, shells, tails, streams, asymmetric haloes and plumes, and double nuclei in bins of stellar mass and photometric redshift. The error bars along the x-axis represent the extent of the stellar mass bins, while the colours indicate the redshift bins. The full sample is binned equally in stellar mass such that each mass bin contains the same number of galaxies but individual mass/redshift bins have varying numbers of galaxies. Each point shows the tidal feature fraction for the corresponding bin, calculated using Equation~\ref{eq:tf_frac_type}, while the upper and lower error bars show the upper and lower bounds for each bin.}
    \label{fig:Feat_type_bars}
\end{figure*}

In this section we examine the distribution of each type of tidal feature with respect to their assigned model classifier scores and the stellar mass, redshift, and colour of their host galaxy. We first investigate whether there exists a relationship between tidal feature types and classifier scores, to determine whether specific tidal feature types are more likely to be assigned higher classifier scores. Fig.~\ref{fig:feat_type_scores} shows the distribution of classifier scores for each tidal feature category. All feature types apart from asymmetric haloes/plumes exhibit a similar trend, peaking in the highest classifier score bin with 30-40 per cent of each type in this bin, declining sharply until classifier scores $\sim0.7$ and more steadily until classifier scores $\sim0.3$. Only $\sim20$ per cent  of plumes and asymmetric haloes are in the highest classifier score bin and instead their distribution peaks for classifier scores 0.8~-~0.9, with $\sim30$ per cent of plumes and asymmetric haloes appearing in this bin. This indicates that the model finds it harder to confidently label galaxies with plumes and asymmetric haloes as tidally disturbed, potentially due to the broader nature of this category. However, this effect does not impact our analysis as we visually classified all galaxies with classifier scores greater than $\sim0.35$, far beyond the point where plumes were assigned high classifier scores less often than other features.

Table~\ref{tab:feat_types} shows the fraction and mean stellar mass, photometric redshift, and $g-i$ colour of host galaxies for 7 categories: galaxies without tidal features, galaxies with any type of feature, and galaxies with shells, tails, streams, asymmetric haloes or plumes, and double nuclei. To calculate the fractions of specific types of feature we use an adjusted version of the method described in Section~\ref{sec:detect}. The calculation of the lower bound remains the same, but when calculating the middle point and upper bound we take into account the prevalence of each type of feature relative to all features present. This is to ensure that the `extra' amount of tidal features added from galaxies not visually classified, when calculating the middle point or upper bound, is not duplicated across multiple tidal feature classes. For example, when considering the upper bound calculation, which assumes a stable level of 8.27 per cent of tidal features for galaxies not visually classified, it would not make sense for non-visually classified galaxies to have an extra 8.27 per cent of both shells and streams. Hence, we can assume that the level of a specific tidal feature class in non-visually classified galaxies is proportional to the prevalence of that tidal feature class relative to the overall number of tidal features in visually classified galaxies. For example, if we consider shells, Equation~\ref{eq:tf_frac} becomes:

\begin{equation}
\label{eq:tf_frac_type}
    f_{\rm{shell}} = \frac{N_{\rm{shell}}}{N_{\rm{gal}}} = \frac{N_{\rm{shell, vis}}}{N_{\rm{gal}}} + \left(\frac{N_{\rm{tidal, est}}}{N_{\rm{gal}}} \times \frac{N_{\rm{shell, vis}}}{N_{\rm{tidal, vis}}}\right)
\end{equation}
where $N_{\rm{gal}}$ is the total number of galaxies in our sample, $N_{\rm{shell, vis}}$ is the number of visually-classified galaxies with shells, and $N_{\rm{tidal, vis}}$ is the number of visually-classified galaxies with tidal features. $N_{\rm{tidal, est}}$ is the estimated number of tidal features in galaxies that were not visually classified, which varies depending on whether we are calculating the lower or upper bound, or the middle estimate, as explained in Section~\ref{sec:detect}.

In our sample, shells are the least common type of tidal feature, and, on average, appear around hosts with slightly higher stellar mass, although this is not statistically significant (Table~\ref{tab:feat_types}). Asymmetric haloes and plumes are the most common type of tidal feature, followed by double nuclei. Although tails and streams can be hard to distinguish visually, the properties of their host galaxies do vary slightly, with hosts of tails having very slightly lower stellar masses than hosts of streams.

We explore the relationship between the occurrence of specific tidal feature types, stellar mass, and photometric redshift in more detail in Fig.~\ref{fig:Feat_type_bars}. This figure shows the fraction of all features, shells, tails, streams, asymmetric haloes or plumes, and double nuclei present in bins of stellar mass and redshift. We can see that while all types of features follow the same general trend, namely being more prevalent in higher stellar mass and lower redshift galaxies, there are slight variations for different feature types. For asymmetric haloes and plumes, which have the highest prevalence of any feature type across the full stellar mass and redshift range, we see the same relationship with stellar mass and redshift as for the all features panel. Asymmetric haloes and plumes are still observed in lower stellar mass, higher redshift bins, where other features are not. This could be attributed to the decreased visibility of fine detail at higher redshifts and lower masses, meaning that the distinct shape of the tidal feature is less identifiable, causing them to fall more frequently into the asymmetric halo and plume category.

Shells consistently have the lowest fractions across all stellar mass and redshift bins. At redshifts $z\geq0.2$, shells only appear at stellar masses $\log_{10}(M_{\star}/\mathrm{M}_{\odot})\geq10.75$, and do not appear at all at $z\geq0.3$. Similarly, at redshifts $z\geq0.3$, streams are only present at stellar masses $\log_{10}(M_{\star}/\mathrm{M}_{\odot})\geq10.75$. Interestingly, for redshifts $z<0.15$, the rate of tails does not show a strong decline with decreasing stellar mass, staying constant for stellar masses $9.50\leq\log_{10}(M_{\star}/\mathrm{M}_{\odot})<10.49$. For stellar masses $\log_{10}(M_{\star}/\mathrm{M}_{\odot})\geq10.75$, double nuclei appear at higher rates in galaxies at $0.15\leq$~$z<0.2$ than at $z<0.15$. This is likely a consequence of galaxy sizes in the cutouts, as high mass, low redshift galaxies would occupy larger areas in our images, causing interacting secondary galaxies to be outside the bounds of the cutouts, as also seen in \citet{Desmons2023GAMA}.

Lastly, since galaxies were only assigned to the double nucleus category if they also possessed another type of tidal feature, we investigate whether double nuclei are more or less likely to be associated with certain classes of tidal features. We find that $40\pm{1}$ per cent of all galaxies with tidal features also exhibit double nuclei. Hosts of tail, streams, and asymmetric haloes and plumes share similar rates of double nuclei of $36\pm{3}$ per cent, $36\pm{4}$ per cent, and $39\pm{1}$ per cent, respectively. Shell hosts, on the other hand, have significantly lower rates of double nuclei, with only $18^{+6}_{-4}$ per cent of them having a visibly interacting secondary galaxy visible. 

\section{Discussion}
\label{sec:disc}
We have used a combination of machine learning and visual classification to identify and classify tidal features in a sample of 34,331 HSC-SSP galaxies with photometric redshifts $z\leq0.4$ and stellar masses $\log_{10}(M_{\star}/\mathrm{M}_{\odot})\geq9.5$. We calculated the tidal feature fraction for the full sample, and examined its dependence on galaxy stellar mass, photometric redshift, and $g-i$ colour, and cluster/group halo mass. We classified tidal features into five classes, and investigated whether distinct trends existed between these feature types and the properties of their host galaxies. In this section we compare these results to those obtained from previous analyses.

\subsection{Tidal feature fraction}
\label{sec:disc_TF_frac}
Using the method outlined in Section~\ref{sec:detect} we calculated a tidal feature fraction $f=0.06~(0.05,~0.11)$ for our sample of 34,331 galaxies, where the numbers in parentheses indicate the lower and upper bounds on the fraction. Table~\ref{tab:TF_frac_lit} in Appendix~\ref{sec:app_TF_lit} provides a comprehensive list of past works in which tidal feature fractions were calculated, along with the available properties of the datasets used to calculate these tidal feature fractions. Large variations exist for tidal feature fractions in the literature, ranging from $f=0.05$ (e.g. \citealt{Adams2012EnvDep}) to $f=0.73$ (e.g. \citealt{Tal2009EllipGalTidalFeat}). These variations can in large part be attributed to the properties of the datasets, namely the depth reached by the imaging survey used to classify the tidal features, and the range of stellar mass and redshift covered by the sample. To conduct fair comparisons between works, one must take into account these variables. Many past works limited their study to massive, red, early-type galaxies (ETGs) (e.g. \citealt{Tal2009EllipGalTidalFeat, Bilek2020MATLASTidalFeat, Huang2022HSCTidalFeatETG, Giri2023ETGRemnants, Rutherford2024MNRAS_SAMI_TF, Yoon2024A_ETG_rot, Yoon2024B_ETG_env}) which are the types of galaxies found to be the most common hosts of tidal features (e.g. \citealt{Atkinson2013CFHTLSTidal, Desmons2023GAMA}). Additionally, many of these works focused on very nearby ($z<0.05$) galaxies for which tidal features are more likely to be observable. As a result, they found tidal feature fractions significantly higher ($f=0.20$ to $f=0.73$) than those calculated in this analysis and are not directly comparable due to their specific sample selection criteria. \citet{Adams2012EnvDep} also focused specifically on tidal features around ETGs located only in cluster environments. The work conducted by \citet{Skryabina2024MNRAS_TF_discs_SDSS} is also not comparable as their sample was limited to edge-on disc galaxies. The work of \citet{Sola2025TF_env} focused on very nearby galaxies ($z\lesssim0.1$), making comparison with their obtained tidal feature fraction difficult due to the lack of galaxies at redshifts this low in our sample.

\citet{Atkinson2013CFHTLSTidal}, \citet{Desmons2023GAMA}, and \citet{Tanaka2023GalCruise} studied broader samples but only at $z<0.2$. To enable comparison with these works we repeat our tidal feature fraction calculation considering the same upper limit on redshift and find $f=0.12~(0.11,~0.14)$, where the numbers in parentheses indicate the lower and upper bound on the fraction. Our classification scheme aligns most with the confidence level 3 in \citet{Atkinson2013CFHTLSTidal}. The likely reason our tidal feature fraction is lower than their confidence level 3 fraction ($f_{\rm{Atkinson, C3}}=0.18\pm0.01)$ is due to the brightness limit of their sample. \citet{Atkinson2013CFHTLSTidal} only considered luminous galaxies with a maximum $r'$-band magnitude $r'=17~\rm{mag}$ around which tidal features are more easily visible, whereas we impose no luminosity limits on our sample which contains galaxies as faint as $r=21.7~\rm{mag}$. If we apply an $r$-band magnitude limit $r<17$ mag in addition to the $z<0.2$ limit, we obtain a tidal feature fraction $f\sim0.20$ which is in good agreement with the \citet{Atkinson2013CFHTLSTidal} confidence level 3 tidal feature fraction.

Although not mentioned in the paper, \citet{Desmons2023GAMA} did assign confidence levels ranging from 0 to 3 to their galaxy classifications and calculated their tidal feature fraction using all classifications with confidence greater than 0. We find that our classification methodology in this work aligns more closely with the level 3 confidence category from \citet{Desmons2023GAMA}. Considering only classifications with confidence scores of 3 in \citet{Desmons2023GAMA}, their tidal feature fraction becomes $f=0.11\pm0.01$ which is in agreement with that found here.

In their citizen science project, GALAXY CRUISE, \citet{Tanaka2023GalCruise} obtained tidal feature classifications for $\sim$20,700 HSC-SSP galaxies and reported a tidal feature fraction $f=0.128\pm0.002$. If we replicate their sample selection by imposing a redshift cut $z<0.2$ and $z$-band magnitude cut $z<17$~mag to our sample we obtain a significantly higher tidal feature fraction $f=0.207(0.205,~0.209)$. One possible explanation for this is that most of the images in their sample (>95 per cent) were sourced from the Wide layer of the HSC-SSP and therefore shallower than those we use from the D/UD layers. This likely impacted the visibility of tidal features. \citet{Tanaka2023GalCruise} also note that their calculated tidal feature fraction is dependent on the threshold chosen to assign galaxies their classification. The fraction $f=0.128\pm0.002$ reported in \citet{Tanaka2023GalCruise} was obtained using a conservative threshold $P>0.79$ but they note that decreasing the threshold to $P>0.5$ would increase the tidal feature fraction to $f\sim0.36$, meaning that there exists a threshold between these where our results would be in agreement.

Comparison with the work of \citet{Hood2018RESOLVETidalFeat} is made difficult due to their sample being limited to very nearby galaxies ($z\sim0.025$) and the lack of galaxies at redshifts this low in our sample. When considering only the low-redshift end of our sample ($z<0.1$) we find a tidal feature fraction $f\sim0.20$, slightly higher than the fraction $f=0.17\pm0.01$ obtained by \citet{Hood2018RESOLVETidalFeat}. This can likely be attributed to differences in the mass range between the samples. \citet{Hood2018RESOLVETidalFeat} used data sourced from the RESOLVE survey which contains galaxies down to stellar masses $M_{\star}\sim10^{8}~\rm{M_{\odot}}$ whereas we only consider galaxies with $M_{\star}\geq10^{9.5}~\rm{M_{\odot}}$.

Perhaps the most relevant observational work with which to compare our results is the analysis performed by \citet{KadoFong2018HSCTidalFeat}. They classified tidal features in $\sim21,000$ HSC-SSP Wide galaxies using a combination of visual classification and an automated method which separates high-spatial frequency features (e.g. tidal features) from low spatial frequency features (e.g. host galaxy light). In their classification, \citet{KadoFong2018HSCTidalFeat} only considered streams and shells and found a tidal feature fraction $f\sim0.06$. For accurate comparison, we recalculate our fraction using the same tidal feature categories (however, based on the description of their stream category, we combine our stream and tail categories) and find a tidal feature fraction $f=0.02~(0.01,~0.03)$. While we might expect a higher tidal feature fraction than that found by \citet{KadoFong2018HSCTidalFeat} due the higher redshift limit ($z=0.45$) of their sample, the lower fraction we find could be explained by differences in sample galaxy brightness and stellar mass distributions. Not only were the galaxies in the \citet{KadoFong2018HSCTidalFeat} sample required to have spectroscopic redshifts from SDSS, causing them to be significantly brighter on average than those in our sample, but their sample stellar mass distribution had a higher concentration of high stellar mass galaxies. Their mean stellar mass of $\log_{10}(M_{\star}/\mathrm{M}_{\odot})=11.19$ is significantly higher than the mean stellar mass of our sample of $\log_{10}(M_{\star}/\mathrm{M}_{\odot})=10.38$. Their use of the larger area of the HSC-SSP Wide layer also likely meant their sample contained more of the rarer brightest galaxies, while missing the fainter galaxies detected in the smaller volume D/UD layers.

In addition to comparison with works based on observations, we can also compare our tidal feature fraction to those obtained from simulations. \citet{Khalid2024TF_Sims} analysed the prevalence and distribution of tidal features in four cosmological hydrodynamical simulations by producing LSST-like mock images. At their highest confidence level (the level which lines up best with our classification criteria) they found a tidal feature fraction $f=0.26\pm0.07$, averaged across the four simulations. All the mock images in the \citet{Khalid2024TF_Sims} sample were produced such that the galaxies were placed very nearby, at a redshift of $z=0.025$. This poses some challenges for direct comparison with our work as we only have a very small sample of galaxies this nearby ($<30$ galaxies). If we consider only the galaxies at redshifts $z<0.1$ in our sample we find a tidal feature fraction $f\sim0.20$, which is in agreement with the \citet{Khalid2024TF_Sims} fraction within 1$\sigma$. The difference between the fractions can likely be explained by the different sample redshifts.

\subsection{Dependence on stellar mass and redshift}
\label{sec:disc_mass_z_col}
In Section~\ref{sec:Mass_z_dep} we explored the relationship between tidal feature fraction and galaxy stellar mass and redshift. We found a positive relationship between stellar mass and tidal feature fraction. When exploring the dependence of the tidal feature fraction on galaxy photometric redshift, we found a negative relationship between tidal feature fraction and galaxy redshift. When exploring the dependence of the tidal feature fraction on both stellar mass and redshift simultaneously, we found that for any given redshift (stellar mass) bin, the tidal feature fraction peaked in the highest stellar mass (lowest redshift) bin. We also found that above redshift $z=0.2$ the dependence of the tidal feature fraction on stellar mass disappeared, and the tidal feature fraction remained approximately constant at $f\sim0.03$ in all stellar mass bins, apart from the highest mass bin $\log_{10}(M_{\star}/\mathrm{M}_{\odot})\geq10.75$.

Qualitatively, the relationship we observe between galaxy stellar mass and tidal feature fractions is in agreement with many works in the literature. The observation-based works of \citet{KadoFong2018HSCTidalFeat, Bilek2020MATLASTidalFeat, Huang2022HSCTidalFeatETG, Desmons2023GAMA, Rutherford2024MNRAS_SAMI_TF, Yoon2024A_ETG_rot}; and \citet{Sola2025TF_env} all find an increasing relationship between tidal feature occurrence and stellar mass. Simulation-based works such as \citet{Pop2018ShellsIllustris}, \citet{Khalid2024TF_Sims}, and \citet{Valenzuela2024_TF_kins_form_hist} also found an increasing trend between stellar mass and the tidal feature fraction. \citet{Bilek2020MATLASTidalFeat} hypothesized that this relationship with tidal feature fraction and stellar mass is a consequence of more massive galaxies having stronger gravitational attraction and tidal forces, causing an increased rate of tidal features around these more massive objects. 

Many of the works listed in Table~\ref{tab:TF_frac_lit} were conducted on very nearby galaxies, and there are limited works studying the relationship between tidal feature fraction and galaxy redshift. The study conducted by \citet{KadoFong2018HSCTidalFeat} considers the prevalence of shells and streams in HSC-SSP Wide layer galaxies up to redshift $z=0.45$. For any given stellar mass bin, they found that the tidal feature fraction decreased with redshift, and above redshifts $z\sim0.25$ detection became difficult and features were almost only visible around host galaxies in the highest stellar mass bin, $\log_{10}(M_{\star}/\mathrm{M}_{\odot})>10.11$. This is in good agreement with our findings, where we found that above redshift $z\sim0.2$, the tidal feature fraction remained approximately constant at very low levels in all stellar mass bins apart from the highest mass bin. \citet{Tanaka2023GalCruise} also considered the effect of redshift on the rate of interacting galaxies in HSC-SSP. Although their study only considered galaxies out to $z=0.2$, they still found a decreasing fraction of visibly interacting galaxies with increasing redshift. Both of these works associate this trend with redshift to the challenges associated with the observability of tidal features at higher redshifts. This observability limitation due to redshift is confirmed by \citet{Martin2022TidalFeatMockIm} who found a factor of $\sim3$ decrease in the number of tidal features detected in identical samples of simulated galaxies placed at $z=0.05$ and $z=0.4$, for a surface brightness limit comparable to that of the HSC-SSP.

\subsection{Dependence on halo mass}
\label{sec:disc_halo_mass}
We investigated whether a relationship exists between tidal features and the environments occupied by their hosts by  examining the distribution of tidal features with respect to halo mass for galaxies in GAMA groups and CAMIRA clusters. We found a slightly higher rate of tidal features in groups with $12.0<\log_{10}(M_{200}/\mathrm{M}_{\odot})<14.0$ than in field galaxies, and a decreasing trend in the tidal feature fraction with increasing halo mass. This suggests that mergers occuring in group environments either form more tidal features, or tidal features in these environments remain observable longer, than those formed from mergers occurring in the field or in denser, cluster environments. We then investigated whether satellite and central galaxies exhibited varying tidal feature trends with halo mass, verifying that the trends we observed could not be accounted for by the stellar mass and redshift distributions present in the bins. We found that the trend of decreasing tidal feature fraction with increasing halo mass present in satellite galaxies could be explained by stellar mass and redshift distribution of those galaxies. However, the stellar mass and redshift distribution could not fully account for central galaxies having significantly higher tidal feature fractions. Our findings suggest that central galaxies in groups and clusters host more tidal features, or host tidal features that remain observable longer, particularly central galaxies in systems with halo masses $\log_{10}(M_{200}/\mathrm{M}_{\odot})>14.1$.

Our findings concerning the relationship between tidal feature incidence and environment are in excellent agreement with the literature. Many works, both observational (e.g. \citealt{Tal2009EllipGalTidalFeat, Adams2012EnvDep, Omori2023ML_TF_env, Tanaka2023GalCruise, Yoon2024B_ETG_env}) and simulation-based (e.g. \citealt{Khalid2024TF_Sims}) found a decrease in tidal feature occurrence with increasing environment density or increasing halo mass. In their analysis of mock images from simulations, \citet{Khalid2024TF_Sims} found that the tidal feature fraction peaked for mergers occurring in systems with halo masses $12.5\lesssim\log_{10}(M_{200}/\mathrm{M}_{\odot})\lesssim13.0$ which is in excellent agreement with our findings. In agreement with the trends observed in our sample, they also found that central galaxies in clusters had significantly higher tidal feature fractions than satellite galaxies, and that, unlike for satellite galaxies, the tidal feature fractions for centrals did not decrease with increasing halo mass but increased instead. Contrary to our observations however, \citet{Khalid2024TF_Sims} found that the trend between tidal features and halo mass for central galaxies could be accounted for by the stellar mass distribution of their galaxies, whereas the relationship of decreasing tidal feature fraction with increasing halo mass for satellite galaxies could not be accounted for by the stellar mass distribution of their galaxies. However, the trends observed here and by \citet{Khalid2024TF_Sims} are subtle in a statistical sense, and the addition of larger, spectroscopically-confirmed samples would help to disentangle this question.

\subsection{Individual feature types and host galaxy properties}
\label{sec:disc_feat_types}
In Section~\ref{sec:type_dep} we examined the prevalence of each type of feature, and the relationship between individual types of tidal features and the properties of their host galaxies, such as colour, stellar mass, and photometric redshift. In this section we compare these results with the trends found in the literature. 

Out of our tidal feature categories, we found shells to be the least common type of feature, followed by streams, tails, double nuclei, and finally asymmetric haloes and plumes being the most common. We also found that hosts of shells had the highest mean stellar masses ($\log_{10}(M_{\star}/\mathrm{M}_{\odot})=10.75\pm0.49$), and hosts of streams had higher mean stellar masses ($\log_{10}(M_{\star}/\mathrm{M}_{\odot})=10.64\pm0.41$) than hosts of tails ($\log_{10}(M_{\star}/\mathrm{M}_{\odot})=10.55\pm0.45$). Although these differences are systematic rather than statistically significant, the direction of the findings are in excellent agreement with the works of \citet{KadoFong2018HSCTidalFeat}, \citet{Desmons2023GAMA}, \citet{Khalid2024TF_Sims}, and \citet{Gordon2024DecalsTFCNN}. Conversely, in their classification of galaxies from the hydrodynamical cosmological simulation Magneticum, \citet{Valenzuela2024_TF_kins_form_hist} found similar fractions of shells and tails, and a significantly higher fraction of streams when considering $z\sim0$ and $\log_{10}(M_{\star}/\mathrm{M}_{\odot})>11$ galaxies. Similar fractions of shells, streams, tails are found in the samples of \citet{Atkinson2013CFHTLSTidal} and \citet{Bilek2020MATLASTidalFeat}, however these samples were composed of massive, luminous galaxies and are therefore difficult to compare.

We also found that the incidence rate of tails at low redshifts did not exhibit a strong relationship with stellar mass, which is in good agreement with \citet{Atkinson2013CFHTLSTidal}, who found a lack of relationship between stellar mass and arms (their classification category resembling our tail category).

Our finding that shells were associated with double nuclei significantly less often than other types of features is in line with the idea that they are formed though mergers that cause the destruction of satellites on near radial orbits \citep{Hendel2015TidalDebOrbit}.

In their analysis of tidal feature detection as a function of limiting surface brightness, \citet{Martin2022TidalFeatMockIm} found that, at $z=0.05$, the increase of the shell fraction with increasing stellar mass persisted over all their investigated surface brightness limits. However, the relationship between stellar mass and the fractions of streams and tails depended on limiting surface brightnesses and the fraction of streams and tails stabilized to $\sim0.4$ across the entire stellar mass range $9.25\leq\log_{10}(M_{\star}/\mathrm{M}_{\odot})<11.25$ at fainter limiting surface brightnesses. This indicates that the prevalence of streams and tails may be independent of host stellar mass, but instead be subject to visibility effects imposed by the limiting surface brightness of observational surveys. This may explain why we observe slightly higher rates of tails than streams, as tails are composed of material expelled from the host galaxy, and may therefore be brighter than streams, which are formed from stellar material pulled out of lower mass companions. If this is the case, the findings of \citet{Martin2022TidalFeatMockIm} also explain our lack of observed relationship between tail incidence rate and stellar mass at $z<0.15$, as the brighter nature of tails would be less affected by the surface brightness limit of our images, particularly at these low redshifts.

\subsection{Broader implications}
\label{sec:merger_rates}
We have investigated how the fraction of galaxies hosting tidal features varies with stellar mass, redshift, and environment, but another key reason for understanding tidal features is to understand how often galaxies undergo mergers. Using our calculated tidal feature fraction and the observability time-scale of tidal features we calculate a rough estimate of the merger rate as follows:
\begin{equation}
    \Re_{\rm{merge}} = \frac{f_{\rm{tidal}}}{t_{\rm{obs}}}
\end{equation}
Here $\Re_{\rm{merge}}$ is the number of mergers per galaxy per Gyr, $f_{\rm{tidal}}$ is the tidal feature fraction, and $t_{\rm{obs}}$ is the length of time for which tidal features remain observable in Gyr.  From their study of HSC-SSP galaxies, \citet{Huang2022HSCTidalFeatETG} estimated that tidal features remain observable for $\sim3~\rm{Gyr}$. \citet{Mancillas2019ETGS_merger_hist} estimated the survival time of various tidal feature types using a cosmological hydrodynamical simulation and found survival times of $\sim2~\rm{Gyr}$, $\sim3~\rm{Gyr}$, and $\sim4~\rm{Gyr}$ for tails, streams, and shells, respectively. Based on these works, we estimate a value of $t_{\rm{obs}}=3~\rm{Gyr}$ for the observability time-scale of tidal features. We calculate the tidal feature fraction, setting a redshift limit $z<0.25$, resulting in a subsample of 14,782 galaxies. The redshift limit ensures we are considering a volume-limited subsample, and limits the impact of low tidal feature observability effects at higher redshifts. Using this subsample, we find a tidal feature fraction $f_{\rm{tidal}}=0.08_{-0.01}^{+0.05}$ and merger rate $\Re_{\rm{merge}}=0.029_{-0.003}^{+0.012}~\rm{Gyr}^{-1}$ where the uncertainties are based on the lower and upper bounds of the tidal feature fraction. Calculations of the merger rate in the literature based on observational data are most commonly obtained by calculating close pair fractions and the merging time-scale of close pairs. Using close pairs in HSC-SSP data, \citet{Huang2022HSCTidalFeatETG} calculated a merger rate of $\Re_{\rm{merge}}=0.086~\rm{Gyr}^{-1}$ for galaxies at $z<0.15$ with stellar masses $\log_{10}(M_{\star}/\mathrm{M}_{\odot})>11$. Both \citet{Mundy2017GAMA_merger_rate} and \citet{Conselice2022merger_rates} calculated the merger rate for galaxies at $z<0.2$ with stellar masses $\log_{10}(M_{\star}/\mathrm{M}_{\odot})>11$ using GAMA close pairs and found lower merger rates $\Re_{\rm{merge}}\sim0.03~\rm{Gyr}^{-1}$ and $\Re_{\rm{merge}}\sim0.02~\rm{Gyr}^{-1}$, respectively. If we further limit our subsample to galaxies with $\log_{10}(M_{\star}/\mathrm{M}_{\odot})>11$, a subsample of 729 galaxies, we find a merger rate $\Re_{\rm{merge}}=0.072_{-0.002}^{+0.004}~\rm{Gyr}^{-1}$, a value closer to that found by \citet{Huang2022HSCTidalFeatETG}. \citet{Rodriguez-Gomez2015Illustris_merger_rate} calculate merger rates as a function of stellar mass using the Illustris cosmological hydrodynamical simulation and at the median stellar mass of our subsample, $\log_{10}(M_{\star}/\mathrm{M}_{\odot})=10.3$ they find $\Re_{\rm{merge}}\sim0.04~\rm{Gyr}^{-1}$ at $z\sim0.1$. At this same stellar mass and redshift \citet{O'Leary2021DMSim_merger_rates} find a merger rate $\Re_{\rm{merge}}\sim0.02~\rm{Gyr}^{-1}$ using the EMERGE N-body simulation. Although our merger rate was calculated using a subsample with a slightly higher median redshift ($z=0.2$), our calculated merger rate agrees reasonably well with the rates calculated based on these simulations. 

Better estimation of the merger rate from tidal features could be achieved by considering the effects of stellar mass and environment on the longevity and visibility of tidal features, and taking these different time-scales into consideration when calculating the merger rate would also likely lead to more robust estimates. Additionally, taking into account the effect of redshift on the visibility of tidal features, as explored by \citet{Martin2022TidalFeatMockIm}, and using upcoming deeper imaging surveys like LSST to limit the impact of low visibility at higher redshifts would also help to further constrain the merger rates.

\section{Conclusions}
\label{sec:conc}
In this work we have used a combination of self-supervised machine learning and visual classification to identify tidal features in a sample of 34,331 galaxies with stellar masses $\rm{log}_{10}(M_{\star}/\rm{M}_{\odot})\geq9.5$ and redshift $z\leq0.4$, drawn from the Hyper Suprime-Cam Subaru Strategic Program optical imaging survey. We assembled the largest sample to-date of 1646 galaxies with confirmed tidal features, and analysed how the incidences of tidal features and the various classes of tidal features vary with host galaxy stellar mass, photometric redshift, colour, and halo mass.
\begin{enumerate}
    \item We found a tidal feature fraction $f=0.06$ with lower and upper bounds $f=0.05$ and $f=0.11$, respectively.
    \item We found an increasing relationship between tidal feature fraction and host galaxy stellar mass, and a decreasing relationship between tidal feature fraction and redshift (Fig.~\ref{fig:Mass_z_box_plots}).
    \item By analysing the relationship between the tidal feature fraction and the halo mass of galaxies, we found evidence suggesting that mergers occurring in group environments with $12.0<\log_{10}(M_{200}/\mathrm{M}_{\odot})<14.0$ either form more tidal features, or tidal features in these environments remain observable longer, than those formed from mergers occurring in the field or in denser, cluster environments (Fig.~\ref{fig:Halo_mass_box_plots}).
    \item We also found that the central galaxies of groups and clusters exhibit higher rates of tidal features than satellite galaxies or than those expected given their stellar masses and redshifts (Fig.~\ref{fig:Centrals_sats_box_plots}).
    \item Out of our individual tidal feature classes, we found shells to be the least common type of feature, followed by streams, tails, double nuclei, and asymmetric haloes/plumes. Although not statistically significant, we found that hosts of shells had the highest mean stellar masses, and hosts of streams had higher mean stellar masses than hosts of tails. We also found a lack of relationship between stellar mass and the incidence rate of tails at low redshifts, suggesting that the observed relationship between tails and stellar mass at higher redshifts may not be intrinsic but rather due to observability limitations. Our finding that shells were associated with double nuclei significantly less often than other types of features is in line with the idea that they are formed though mergers that cause the destruction of satellites on near radial orbits (Fig.~\ref{fig:Feat_type_bars}).
    \item We also estimated the merger rate based on tidal feature observability time-scales, and found that on average, galaxies in our sample undergo $0.029$ mergers per Gyr.
\end{enumerate}

 We found good agreement between the trends we observe and the results obtained from purely visual or other automated methods, confirming the validity of our methodology and that using machine learning can drastically reduce the workload of visual classifiers, having needed to visually classify less than 30 per cent of our sample. Such methods will be instrumental in classifying the millions of galaxies observed by upcoming large optical imaging surveys such as the Vera C. Rubin Observatory’s Legacy Survey of Space and Time. We make our catalogue of galaxies and their tidal feature classifications publicly available on Zenodo \citep{Desmons2025ZenodoCat}.

\section*{Acknowledgements}
\label{sec:acknowledge}
We thank the anonymous reviewer for their constructive comments which have improved the quality of this paper. We thank Edward Taylor for helpful discussions regarding colour measurements and stellar mass calculations. This research made use of the ``K-corrections calculator'' service available at \url{http://kcor.sai.msu.ru/}. We acknowledge funding support from LSST Corporation Enabling Science grant LSSTC 2021-5. SB acknowledges funding support from the Australian Research Council through a Discovery Project DP190101943. The Hyper Suprime-Cam (HSC) collaboration includes the astronomical communities of Japan and Taiwan, and Princeton University. The HSC instrumentation and software were developed by the National Astronomical Observatory of Japan (NAOJ), the Kavli Institute for the Physics and Mathematics of the Universe (Kavli IPMU), the University of Tokyo, the High Energy Accelerator Research Organization (KEK), the Academia Sinica Institute for Astronomy and Astrophysics in Taiwan (ASIAA), and Princeton University. Funding was contributed by the FIRST program from the Japanese Cabinet Office, the Ministry of Education, Culture, Sports, Science and Technology (MEXT), the Japan Society for the Promotion of Science (JSPS), Japan Science and Technology Agency (JST), the Toray Science Foundation, NAOJ, Kavli IPMU, KEK, ASIAA, and Princeton University. 
This paper makes use of software developed for Vera C. Rubin Observatory. We thank the Rubin Observatory for making their code available as free software at http://pipelines.lsst.io/.
This paper is based on data collected at the Subaru Telescope and retrieved from the HSC data archive system, which is operated by the Subaru Telescope and Astronomy Data Center (ADC) at NAOJ. Data analysis was in part carried out with the cooperation of Center for Computational Astrophysics (CfCA), NAOJ. We are honored and grateful for the opportunity of observing the Universe from Maunakea, which has the cultural, historical and natural significance in Hawaii.

GAMA is a joint European-Australasian project based around a spectroscopic campaign using the Anglo-Australian Telescope. The GAMA input catalogue is based on data taken from the Sloan Digital Sky Survey and the UKIRT Infrared Deep Sky Survey. Complementary imaging of the GAMA regions is being obtained by a number of independent survey programmes including GALEX MIS, VST KiDS, VISTA VIKING, WISE, Herschel-ATLAS, GMRT and ASKAP providing UV to radio coverage. GAMA is funded by the STFC (UK), the ARC (Australia), the AAO, and the participating institutions. The GAMA website is https://www.gama-survey.org/ .

This research used data obtained with the Dark Energy Spectroscopic Instrument (DESI). DESI construction and operations is managed by the Lawrence Berkeley National Laboratory. This material is based upon work supported by the U.S. Department of Energy, Office of Science, Office of High-Energy Physics, under Contract No. DE–AC02–05CH11231, and by the National Energy Research Scientific Computing Center, a DOE Office of Science User Facility under the same contract. Additional support for DESI was provided by the U.S. National Science Foundation (NSF), Division of Astronomical Sciences under Contract No. AST-0950945 to the NSF’s National Optical-Infrared Astronomy Research Laboratory; the Science and Technology Facilities Council of the United Kingdom; the Gordon and Betty Moore Foundation; the Heising-Simons Foundation; the French Alternative Energies and Atomic Energy Commission (CEA); the National Council of Science and Technology of Mexico (CONACYT); the Ministry of Science and Innovation of Spain (MICINN), and by the DESI Member Institutions: www.desi.lbl.gov/collaborating-institutions. The DESI collaboration is honored to be permitted to conduct scientific research on Iolkam Du’ag (Kitt Peak), a mountain with particular significance to the Tohono O’odham Nation. Any opinions, findings, and conclusions or recommendations expressed in this material are those of the author(s) and do not necessarily reflect the views of the U.S. National Science Foundation, the U.S. Department of Energy, or any of the listed funding agencies.\\ 
%%%%%%%%%%%%%%%%%%%%%%%%%%%%%%%%%%%%%%%%%%%%%%%%%%
\section*{Data Availability}
\label{sec:data_avail}
 
 All data used in this work is publicly available at \url{https://hsc-release.mtk.nao.ac.jp/doc/} (HSC-SSP data), \url{https://www.gama-survey.org/} (GAMA data), and \url{https://data.desi.lbl.gov/doc/} (DESI data). The machine learning model and instructions for its use are available on GitHub and can be downloaded from \url{https://github.com/LSSTISSC/Tidalsaurus}.


\bibliographystyle{mnras}
\bibliography{bibs} % if your bibtex file is called example.bib

%%%%%%%%%%%%%%%%%%%%%%%%%%%%%%%%%%%%%%%%%%%%%%%%%%

%%%%%%%%%%%%%%%%% APPENDICES %%%%%%%%%%%%%%%%%%%%%
\appendix

\section{Group and cluster selection}
\label{sec:app_cluster}
\begin{figure}
    \centering
	\includegraphics[width=0.99\columnwidth]{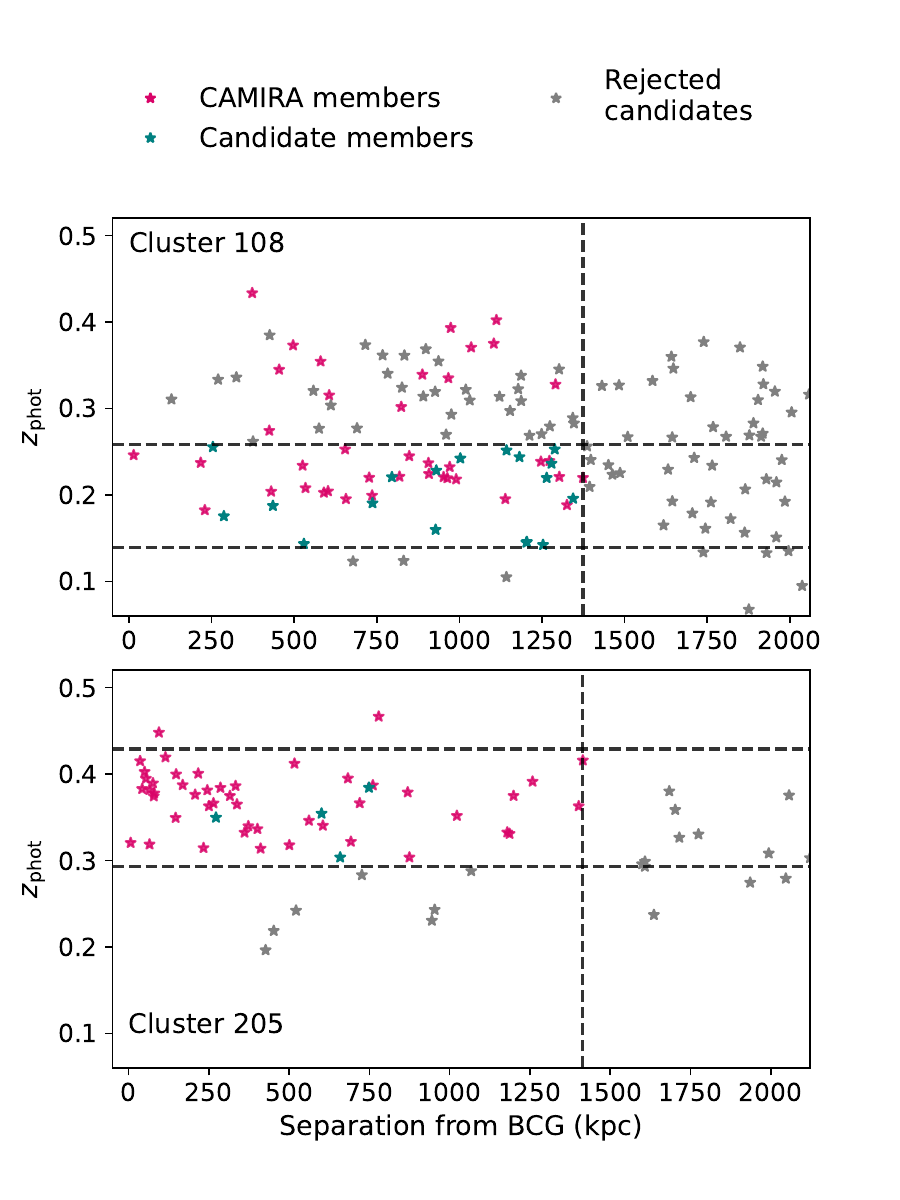}
    \caption{Example of the procedure to identify additional candidate member galaxies for two CAMIRA clusters. The pink shows the confirmed CAMIRA members for each cluster, the teal points show the accepted candidate members for each cluster, and the grey points show the potential candidates rejected either due to being beyond the cut-off in separation or photometric redshift. The vertical dashed line shows the cut-off in projected space, given by the separation between the cluster BCG and the farthest confirmed CAMIRA member. The horizontal dashed lines show the cut-offs in redshift of $\pm2\sigma$ from the BCG redshift, where the $1\sigma$ accuracy of photometric redshifts is $0.05(1+z_{\rm{phot}})$ \citep{Tanaka2018HSCSSP_photoz}.}
    \label{fig:Cand_select}
\end{figure}

This section provides examples on the selection of CAMIRA clusters and GAMA groups described in Section~\ref{sec:Data_clusters} and describes our investigations into the effects of these selections on our results. Fig.~\ref{fig:Cand_select} shows how we identify extra candidate members for the CAMIRA clusters, by considering galaxies in our sample of 34,331 galaxies which are close in projected distance and redshift to CAMIRA cluster members. Table~\ref{tab:over_select_table} lists the single CAMIRA clusters which are composed of multiple GAMA groups and our decisions and justifications regarding whether to use the CAMIRA or GAMA definition of each cluster or group.

\begin{table*}
\caption{CAMIRA clusters composed of multiple GAMA groups, our choice of the CAMIRA or GAMA definition for each system in our final sample, and our justification.}
\label{tab:over_select_table}
\normalsize
\centering
\begin{tabular}{c>
{\centering}m{2.1cm}>
{\centering}m{2.cm}>
{\centering\arraybackslash}m{8.5cm}}
\toprule
 \textbf{CAMIRA Cluster} & \textbf{GAMA Group} & \textbf{Chosen}\\\textbf{Definition} & \textbf{Justification}\\
\midrule
3 & 400013, 400043, 400282 & GAMA & GAMA groups occupy distinct positions in projected separation.
\\
36 & 400049, 400070 & CAMIRA & Fainter CAMIRA members remove clear separation between GAMA groups.
\\
41 & 400219, 400221 & CAMIRA & Fainter CAMIRA members remove clear separation between GAMA groups.
\\
46 & 400039, 400069, 400385 & GAMA & GAMA groups 400069 and 400385 occupy distint positions in projected separation. GAMA group 400039 occupies distinct position in redshift.
\\
56 & 400109, 400145 & GAMA & GAMA groups occupy distinct positions in projected separation and redshift.
\\
44, 45 & 400032, 400143 & GAMA & GAMA groups occupy distinct positions in redshift. \\

\bottomrule
\end{tabular}
\end{table*}

We repeat the analysis presented in Section~\ref{sec:halo_dep}, regarding the relationship between the tidal feature fraction and cluster halo mass and redshift, using a range of subsamples in which our cluster and group selection is varied slightly. This is to ensure that the conclusions we draw are not dependent on the way in which we define our group and cluster samples.

The first two subsamples on which we repeat our analysis pertain to our decision to use the GAMA or CAMIRA groupings for the 105 overlapping members shared between 16 CAMIRA clusters and 23 GAMA groups. In the first of these samples we use only the CAMIRA definition of the 16 clusters (discarding the overlapping GAMA groups), and only the GAMA definitions of the 23 groups in the other sample (discarding the overlapping CAMIRA clusters). Any qualitative conclusions made in Section~\ref{sec:halo_dep} also hold true for these two samples.

To ensure that the challenges associated with member and candidate member identification of CAMIRA clusters, namely the uncertainties associated with photometric redshift, do not affect our conclusions, we also repeat the analysis presented in Section~\ref{sec:halo_dep} considering a smaller sample of CAMIRA clusters for which spectroscpic redshifts are available. Our aim is to ensure that our conclusions concerning the relationship between tidal features and halo mass are not altered when using a subset of CAMIRA members whose memberships can be confirmed spectroscopically. We do this by cross-matching our 1871 CAMIRA members and 1062 candidate members with the DESI EDR spectroscopic data detailed in Section~\ref{sec:data}, obtaining matches for 1056 members and 1149 candidates from 65 distinct clusters. To test the impact of photometric redshifts we study a subset of 17 clusters which have spectroscopic data available for at least 80 per cent of their CAMIRA members. We assess the membership of CAMIRA identifications by considering their spectroscopic velocity relative to the cluster velocity dispersion, which we obtain by using the relationship between halo mass $M_{200}$ and velocity dispersion $\sigma_{\rm{cl}}$ (e.g. \citealt{Finn_HaloMass_vdisp2005}):
\begin{align}
    \left(\frac{\sigma_{\rm{cl}}}{1000~\rm{km~s^{-1}}}\right)^3 = \frac{\sqrt{\Omega_{\Lambda}+\Omega_M(1+z)^3}}{1.2\times10^{15}}\frac{\rm{M}_{\odot}}{M_{200}}
\end{align}
We keep members for which $|\Delta v|~<~3\sigma_{\rm{cl}}$, where $\Delta v~=~c~\times~\Delta z$, $c$ is the speed of light, and $\Delta z$ is the difference between the member's or candidates's spectroscopic redshift and the spectroscopic redshift of the BCG. Using this limit we assemble two versions of the subsample of 17 clusters with >80 per cent spectroscopic completeness. One where we keep only members and candidates with spectroscopic redshifts for whom $|\Delta v|~<~3\sigma_{\rm{cl}}$. The other version involves keeping members and candidates which have spectroscopic redshifts for whom $|\Delta v|~<~3\sigma_{\rm{cl}}$, as well as the members and candidates which only have photometric redshifts. This is to ensure that our conclusions are not biased by the DESI EDR sample containing, on average, brighter galaxies than our parent sample, and missing some of the fainter galaxies. We verify that the conclusions we present in Section~\ref{sec:halo_dep} are maintained when the CAMIRA cluster sample is replaced with either of these two subsets, finding no qualitative changes in the trends observed.

Our final subsample for which we repeat our analysis is constructed to ensure that the differently defined central galaxy in CAMIRA and GAMA does not affect our results. We repeat our analysis by replacing our selection of the GAMA central galaxies with the GAMA BCGs instead. All qualitative results presented in Section~\ref{sec:halo_dep} also hold true when using this definition of GAMA central galaxies.

\section{Tidal feature fractions in the literature}
Table~\ref{tab:TF_frac_lit} provides a comprehensive list of past works in which tidal feature fractions were calculated, along with the available properties of the datasets used to calculate these tidal feature fractions.
\label{sec:app_TF_lit}
\begin{table*}
\caption{Tidal feature fractions in the literature, based on observations. Where available, we list the properties of the sample classified in the work (surface brightness limit, magnitude limit, stellar mass and redshift range), the types of galaxies classified, and the number of objects in the sample. The 'C4', 'C3', 'C2' notations next to tidal feature fractions indicate the fraction calculated for different confidence levels.}
\label{tab:TF_frac_lit}
\normalsize
\centering
\begin{tabular}{p{3.cm}>
{\centering}m{4.cm}>
{\centering}m{1.3cm}>
{\centering}m{3.5cm}>
{\centering}m{1.cm}>
{\centering\arraybackslash}m{2.1cm}}
\toprule
{Paper (survey)} & {Sample properties} & {Galaxy \\ Type} & {Feature Type} & {N$_{\rm{galaxies}}$} & {$f_{\rm{tidal}}$} \\
\midrule
\citet{Yoon2024B_ETG_env} \newline (DESI Legacy Survey) &$\mu_{r}<27.86~\rm{mag}~\rm{arcsec}^{-2}$ $M_{\star}\geq10^{11.2}~\rm{M_{\odot}}$ $0.01<z_{\rm{spec}}<0.04$ & ETGs & Tails, streams, shells & 46 & 0.447 \\
\midrule
\citet{Yoon2024A_ETG_rot} \newline (DESI Legacy Survey) &$\mu_{r}<27.86~\rm{mag}~\rm{arcsec}^{-2}$ $M_{\star}\geq10^{9.65}~\rm{M_{\odot}}$ $z_{\rm{spec}}<0.055$ & ETGs & Tails, streams, shells & 1244 & 0.204 \\
\midrule
\citet{Rutherford2024MNRAS_SAMI_TF} \newline (HSC-SSP Wide) & $\mu_{r}<27.8~\rm{mag}~\rm{arcsec}^{-2}$ $M_{\star}\geq10^{10}~\rm{M_{\odot}}$ $~~~~~z_{\rm{spec}}<0.1~~~~~$ & ETGs & Streams, shells & 411 & $0.31\pm0.02$ \\
\midrule
\citet{Giri2023ETGRemnants} \newline (SDSS Stripe82) & $\mu_{r}<28.5~\rm{mag}~\rm{arcsec}^{-2}$ $z_{\rm{spec}}<0.05$ & ETGs & Tails, streams/plumes, shells, rings, X structures, boxy/discy isophotes, disturbed galaxies & 202 & 0.27 (major) 0.57 (minor) \\
\midrule
\citet{Huang2022HSCTidalFeatETG} \newline (HSC-SSP Wide) & $\mu_{r}<27.8~\rm{mag}~\rm{arcsec}^{-2}$ $M_{\star}\geq10^{11}~\rm{M_{\odot}}$ $0.05<z_{\rm{spec}}<0.15$ & ETGs & -- & 2649 & 0.28 \\
\midrule
\citet{Bilek2020MATLASTidalFeat} \newline (MATLAS) & $\mu_{g}<28.5~\rm{mag}~\rm{arcsec}^{-2}$ $z_{\rm{spec}}<0.01$ & ETGs & Tails, streams, shells, plumes & 177 & $0.41\pm0.06$ \\
\midrule
\citet{Adams2012EnvDep} \newline (MENeaCS) & $M_r<-20$ $\mu_{r}<26.5~\rm{mag}~\rm{arcsec}^{-2}$ $0.04<z<0.15$ & Cluster ETGs & -- & 3551 & $0.053\pm0.004$ \\
\midrule
\citet{Tal2009EllipGalTidalFeat} \newline (OBEY) & $M_B<-20$  $\mu_{V}<27.7~\rm{mag}~\rm{arcsec}^{-2}$ $z_{\rm{spec}}<0.01$ & Ellipticals & Shells, tails, fans, highly disturbed & 55 & 0.73 \\
\midrule
\citet{Skryabina2024MNRAS_TF_discs_SDSS} \newline (HSC-SSP \& \newline DESI Legacy Survey) & $\mu_{r,\rm{HSC}}<29.6~\rm{mag}~\rm{arcsec}^{-2}$ $r_{\rm{HSC}}<28~\rm{mag}$ $r_{\rm{DESI}}<28.4~\rm{mag}$ $~~~z_{\rm{spec}}<0.2~~$ & Edge-on discs & Tails, streams, shells, plumes, bridges, arcs, disc deformations & 838 & 0.058 \\
\midrule
\citet{Sola2025TF_env} \newline (MATLAS, CFIS, \newline VESTIGE, NGVS) &$\mu_{r}\lesssim29~\rm{mag}~\rm{arcsec}^{-2}$ $M_{\star}\gtrsim10^{9.8}~\rm{M_{\odot}}$ ~~~~$z_{\rm{spec}}<0.01$~~~~ & All & Tails, streams, shells & 475 & 0.36 \\
\midrule
\citet{Tanaka2023GalCruise} \newline (HSC-SSP \newline Wide+Deep) &$\mu_{r}<27.8~\rm{mag}~\rm{arcsec}^{-2}$ $~~~~~~z<17~\rm{mag}~~~~~~$ $~~~z_{\rm{spec}}<0.2~~~$ & All & Tails/streams, shells, distorted shapes, rings & 20,686 & $0.128\pm0.002$ \\
\midrule
\citet{Desmons2023GAMA} \newline (HSC-SSP D/UD) & $\mu_{r}<29.8~\rm{mag}~\rm{arcsec}^{-2}$ $10^{9.5}\leq M_{\star}\leq10^{11}~\rm{M_{\odot}}$ $0.04<z_{\rm{spec}}<0.2$ & All & Tails/streams, shells, plumes, double nuclei & 852 & $0.23\pm0.02$ \\
\midrule
\citet{KadoFong2018HSCTidalFeat} \newline (HSC-SSP Wide) & $\mu_{r}<26.4~\rm{mag}~\rm{arcsec}^{-2}$ $0.05<z_{\rm{spec}}<0.45$ & All & Streams/tails, shells & 21,208 & 0.057 \\
\midrule
\citet{Hood2018RESOLVETidalFeat} \newline (RESOLVE) & $\mu_{r}<27.9~\rm{mag}~\rm{arcsec}^{-2}$ $z_{\rm{spec}}<0.025$& All & Narrow (streams, arms), broad (shells, fans, plumes) & 1048 & $0.04\pm0.01$~(C4) $0.17\pm0.01$~(C3) \\
\midrule
\citet{Atkinson2013CFHTLSTidal} \newline (CFHTLS) & $M_{r'}<-19.3$ $\mu_{g}<27.7~\rm{mag}~\rm{arcsec}^{-2}$ $10^{9.5}\leq M_{\star}\leq10^{11.5}~\rm{M_{\odot}}$ $0.04<z_{\rm{spec}}<0.2$ & All & Streams, arms, linear features, fans, shells, plumes & 1781 &  $0.12\pm0.01$~(C4) $0.18\pm0.01$~(C3) $0.25\pm0.01$~(C2)\\
\bottomrule
\end{tabular}
\end{table*}
%%%%%%%%%%%%%%%%%%%%%%%%%%%%%%%%%%%%%%%%%%%%%%%%%%

% Don't change these lines
\bsp	% typesetting comment
\label{lastpage}
\end{document}